# Domain Walls in Supersymmetric Yang–Mills Theories[⋆]

Vadim S. Kaplunovsky,[*]
Jacob Sonnenschein[†] and Shimon Yankielowicz[†]

∗ *Theory Group, Physics Dept., University of Texas, Austin, TX 78712, USA*

† *School of Physics and Astronomy, Tel Aviv University, Ramat Aviv, Tel Aviv 69978, Israel*

## ABSTRACT

We study BPS saturated domain walls in the supersymmetric $SU(2)$ gauge theory. For a theory with a very light adjoint scalar (mass $m \lesssim \Lambda/400$) we use the perturbed $\mathcal{N} = 2$ Seiberg–Witten theory to calculate the actual field configuration of the domain wall. The wall has a sandwich-like five-layer structure of three distinct phases — electric confinement, Coulomb and oblique confinement — separated by two separate transition regions. For larger scalar masses, the three-phase structure disappears and the Seiberg–Witten theory becomes inadequate because of two major problems: First, the higher-derivative interactions between the light fields become relevant and second, both the magnetic monopole condensate and the dyon condensate show up in the same region of space, a phenomenon indescribable in terms of a local field theory. Nevertheless, we argue that the BPS saturated domain wall continues to exist in this regime and give a qualitative description of the scalar and gaugino condensates. Finally, we discuss the domain walls in MQCD and translate the BPS conditions into coupled non-linear differential equations.

---

⋆ Research supported in part by the US–Israeli Binational Science Foundation, the US National Science Foundation (V. K., grant #PHY–95–11632), the Robert A. Welsh Foundation (V. K.).

# 1. Introduction

Domain walls interpolating between degenerate discrete vacua have been a subject of a rather intensive study in recent years.[1−9] In particular, an $\mathcal{N} = 1$ Supersymmetric $SU(N_c)$ Yang–Mills theory has a non-anomalous $Z_{2N_c}$ chiral symmetry which is spontaneously broken down to the $Z_2$ by the expectation value of the gaugino bilinear $\langle \text{tr}\, \lambda^\alpha \lambda_\alpha \rangle$. The SSYM theory thus has $N_c$ degenerate discrete vacua, each characterized by a different value of the chiral gaugino condensate[10]

$$\langle \text{tr}\, \lambda^\alpha \lambda_\alpha \rangle \;=\; 16\pi^2 \Lambda^3 \, e^{2\pi i n/N_c}, \quad n = 1, 2, \ldots, N_c. \tag{1.1}$$

Of special interest are the so-called BPS-saturated domain walls which preserve half of the $\mathcal{N} = 1$ supersymmetry. Fields configurations for such BPS-saturated walls satisfy first-order differential equations, which follow in a straightforward manner from the effective Lagrangian of the theory. Like other BPS-saturated states, the BPS domain wall are more tractable and one may reasonably hope for some exact results for such walls even in a context of a confining strongly interacting theory. Indeed, the tension *i.e.* energy per unit area of a BPS domain wall is exactly determined by the difference between the superpotential values in the two vacua connected by the wall. In the $\mathcal{N} = 1$ SSYM theory, the superpotential — which acts as a central charge for domain walls — is related by the chiral anomaly to the gaugino condensate, so a BPS domain wall has tension[2]

$$\mathcal{T} \;\equiv\; \frac{\text{Energy}}{\text{Area}} \;=\; \frac{N_c}{8\pi^2} \left| \Delta \langle \text{tr}\, \lambda^\alpha \lambda_\alpha \rangle \right|. \tag{1.2}$$

For the $\mathcal{N} = 1$ SQCD theories with $N_f < N_c$, the effective superpotential is so constrained by the twin requirements of the holomorphy and flavor symmetry that one can completely determine its exact form.[10,11,12] Unfortunately, no such constraints apply to the effective Kähler function of the theory which controls the kinetic energies of the fields. The Kähler metric of the field space plays important role in the BPS equations, so the very existence of BPS-saturated domain walls is a non-trivial dynamical question. Thus far, all the investigations of this issue have *assumed* specific Kähler functions, only to find that the answer depends on their assumptions. Furthermore, the singularities of the Kähler metric can lead to additional vacuum states of the theory; indeed, the claims[1–4] that SQCD may have a chirally-invariant vacuum are based on precisely such a Kähler singularity.

The situation is under much better control for the $\mathcal{N}=2$ SQCD where the Kähler metric follows from a holomorphic pre-potential and the entire low-energy effective Lagrangian is completely determined by the Seiberg–Witten theory.[13] In the $\mathcal{N}=1$ terms, the $\mathcal{N}=2$ SQCD has an extra chiral superfield in the adjoint representation of the gauge group. Giving this superfield a mass $m \neq 0$ breaks the supersymmetry down to $\mathcal{N}=1$; in the $m \to \infty$ limit, the adjoint superfield decouples from the low-energy physics and one is left with an effective $\mathcal{N}=1$ SQCD. We have therefore decided to study the BPS-saturated domain walls in the $\mathcal{N}=2$ SQCD perturbed by the adjoint mass $m$. For the sake of simplicity, we focus on the $SU(2)$ SSYM theory without any quarks. An overview of the $SU(2)$ Seiberg–Witten (SW) theory perturbed by the adjoint mass is presented in section 2 of this article; we emphasize the vacuum structure of the theory.

In section 3, we discuss BPS-saturated domain walls in a toy model with a simplified version of the Seiberg–Witten superpotential and a naive *i.e.*, flat Kähler metric. Also the electric and the magnetic charges of the SW theory are replaced with two purely electric charges of an $U(1) \times U(1)$ gauge theory. We write down the BPS equation for a domain-wall field configuration and solve them analytically for a particular value of the mass parameter and numerically otherwise. It turns out that in the small mass limit (analogous to $m \ll \Lambda$), the domain wall's profile develops a sandwich-like five-layer structure. In each of the two outer layers, the fields asymptote to their respective vacuum values and one combination of the two $U(1)$ charges is Higgsed down; this behavior corresponds to the two confining phases of the SW theory characterized by respectively magnetic monopole or dyon condensates. In the middle layer, the toy model is in its Coulomb phase, the $U(1) \times U(1)$ gauge symmetry is essentially unbroken and the modulus field slowly interpolates between its stable-vacuum values; for mass $\neq 0$, the Coulomb phase is thermodynamically unstable in the bulk but exists in a layer of finite thickness inside the domain wall. The two remaining layers contain transition regions between the Coulomb and the appropriate confining phases.

Section 4 of this article is devoted to the analysis of BPS domain walls in Seiberg–Witten theory with a small adjoint mass $m \ll \Lambda$. We argue that in the small mass regime, the SW domain wall has the same five-layer structure as the toy model. We write down the BPS equations for each of the wall's layers and we solve those equations numerically. We find a surprisingly low limit $m \lesssim \Lambda/400$ for the clear five-layer structure of the SW domain wall; beyond this limit, the transition regions take over the Coulomb phase region and overlap each other. Also, the wall becomes too thin to be analyzed in terms of a low-energy *i.e.* long-distance effective theory such as Seiberg–Witten; this problem is discussed in section 5. Nevertheless, we shall argue that the BPS-saturated domain wall exists for any $m$, small or large.

In section 6 we discuss gaugino condensation in the Seiberg–Witten theory. We write down an effective superpotential and analyse the vacuum structure of the theory in terms of both gaugino and scalar condensates. We find the same two stable vacua for all $m \neq 0$ and no trace of any additional chirally invariant vacuum. We also deduce the qualitative behavior of the gaugino condensates inside the domain wall. For finite $m$, there are distinct "photino" and "Wino" condensates and their profiles inside the wall are quite different.

E. Witten[14] has advocated the MQCD approach based on the M–theory. Hopefully, the $\mathcal{N} = 1$ MQCD is in the same universality class as the ordinary $\mathcal{N} = 1$ SQCD. In MQCD, the BPS-saturated domain walls arise naturally and have a geometric interpretation;[16] there is even a geometric interpretation of the QCD string emanating from a probe quark and terminating on the wall.

Witten argues that in the large $N_c$ limit, the BPS wall's tension should scale as $\mathcal{T} \propto N_c$,[16] which makes it hard to interpret the wall as a soliton in the effective low-energy theory since the tension of such a solitonic wall would scale as $\mathcal{T} \propto N_c^2$.[*]

In section 7, we follow Witten's geometric construction of a BPS domain wall in MQCD and write down explicit BPS equations for the wall's geometry. The equations are rather cumbersome, although we managed to simplify them and recast them as a pair of partial differential equations for two complex functions. In the $SU(2)$ case, the two functions are actually real and the BPS equations can be further reduced to a single fourth-order Monge–Ampere equation[17,18] for one real function. Unfortunately, we were unable to find an analytic solution compatible with the domain-wall boundary conditions. We also present simplified equations for the MQCD analogue of the low-mass Seiberg–Witten domain wall; again, we do not have a solution. We hope however that someone will eventually solve our equations; such a solution would be very useful for studying many physical properties of the domain walls in MQCD — and ultimately, in QCD itself.

## 2. Overview of Seiberg–Witten Theory

The strongest evidence for the vacuum structure of the $\mathcal{N} = 1$ SSYM theories comes from the Seiberg–Witten (SW) theory of the $\mathcal{N} = 2$ SSYM.[13] A non-abelian gauge theory with unbroken $\mathcal{N} = 2$ supersymmetry and without any 'quark' hypermultiplets has a continuous family of exactly degenerate vacua. The *moduli space* of these vacua has a complex dimension equal to the rank $r$ of the gauge group. For a generic vacuum, the theory is in the abelian Coulomb phase: The gauge group $G$ is spontaneously broken down to its maximal abelian subgroup $U(1)^r$ and the only massless particles are the $r$ gauge bosons and their $\mathcal{N} = 2$ superpartners. However, the moduli space has singular subspaces where some charged hypermultiplets of particles are also massless.

---

[*] See however [7] for efforts to explain the $\mathcal{T} \propto N_c$ scaling behavior in the field theory framework.

For the sake of simplicity, we focus this article on the $SU(2)$ SSYM theory where the vacuum expectation value $U = \langle \operatorname{tr} \Phi^2 \rangle^\star$ serves as a global complex coordinate of the one-dimensional moduli space of the theory. Thanks to the $\mathcal{N} = 2$ supersymmetry, the Kähler metric for the $U$ field is known exactly as

$$g_{U\overline{U}}(U) = \frac{1}{4\pi} \operatorname{Im}\left( \frac{\partial A_2}{\partial U} \frac{\partial A_1^*}{\partial U^*} \right), \tag{2.1}$$

where $A_1(U)$ and $A_2(U)$ are the two periods of the Seiberg–Witten elliptic curve,

$$A_{1,2}(U) = \frac{\sqrt{2}}{2\pi} \oint_{\text{contour}_{1,2}} \frac{(x-U)\,dx}{\sqrt{x^3 - 2Ux^2 + \Lambda^4 x}}. \tag{2.2}$$

The $U$-dependence of the abelian gauge coupling is given by

$$\frac{2\pi i}{g^2} + \frac{\Theta}{4\pi} \equiv \tau = \frac{dA_2(U)}{dA_1(U)}; \tag{2.3}$$

this analytic function has non-trivial monodromies around its singularities; physically, the monodromies amount to electric-magnetic duality transformations,

$$\tau \to \frac{a\tau + b}{c\tau + d}, \qquad (\mathbf{E} - i\mathbf{B}) \to (c\tau + d)(\mathbf{E} - i\mathbf{B}). \tag{2.4}$$

In duality-invariant terms, the $U$-dependence of the gauge coupling can be summarized as

$$j(\tau) = \frac{(16U^2 - 12\Lambda^2)^3}{\Lambda^8(U^2 - \Lambda^4)}. \tag{2.5}$$

At two points in the moduli scape, namely at $U = \pm\Lambda^2$, a magnetically charged hypermultiplet becomes massless and the gauge coupling (2.3) and the

---

$\star$ The $\Phi$ here is the adjoint multiplet of the gauge symmetry comprised of the scalar $\mathcal{N} = 2$ superpartners of the gauge fields; the trace is taken in the fundamental representation.

Kähler metric (2.1) have logarithmic singularities; mathematically, this corresponds to degeneration of the SW elliptic curve. The hypermultiplet $\mathcal{M}$ which becomes massless at $U = +\Lambda^2$ is a magnetic monopole without electric charge, $(q_e = 0, q_m = 1)$, while at $U = -\Lambda^2$, the massless hypermultiplet is a dyon with charges $(q_e = 1, q_m = 1)$. For $U$ being close to either singularity, the low-energy regime of the SW theory is described by a *local* effective field theory, namely $\mathcal{N} = 2$ SQED with one "electron" — which is actually $\mathcal{M}$ or $\mathcal{D}$ whose charge is rendered purely electric by means of a suitable electric-magnetic duality transformation.

From the $\mathcal{N} = 1$ point of view, the $\mathcal{N} = 2$ SSYM theory is a gauge theory with a chiral supermultiplet $\Phi$ in the adjoint representation of the gauge group, and the second supersymmetry can be softly broken by giving $\Phi$ a non-zero mass. When this happens to the Seiberg–Witten theory, the mass term $m \operatorname{tr}(\Phi^2)$ for the $SU(2)$ triplet $\Phi$ in the superpotential becomes the O'Raighfeartaigh F-term[26] $mU$ for the $U$ modulus (viewed now as an $\mathcal{N} = 1$ chiral superfield). Consequently, the energy density of a generic $\mathcal{N} = 2$ SW vacuum (where $mU$ comprises the entire superpotential of the effective low-energy theory) is lifted to $|m|^2 g^{U\overline{U}} > 0$.

On the other hand, the singular vacua with $U = \pm\Lambda^2$ remain $\mathcal{N} = 1$ supersymmetric vacua: Instead of breaking SUSY, the O'Raighfeartaigh term now causes confinement. Indeed, at the $U = +\Lambda^2$ singularity, the superpotential of the effective low-energy theory is

$$\begin{aligned} W &= mU + A_2(U)\,\mathcal{M}\widetilde{\mathcal{M}} \\ &\approx mU + \frac{\Lambda^2 - U}{i\sqrt{2}\Lambda}\,\mathcal{M}\widetilde{\mathcal{M}} \end{aligned} \quad (2.6)$$

where $\mathcal{M}$ and $\widetilde{\mathcal{M}}$ are the chiral monopole/antimonopole superfields. This superpotential — together with the $U(1)$ gauge superfield magnetically dual to the electric $U(1) \subset SU(2)$ — has a unique supersymmetric vacuum

$$U = +\Lambda^2, \qquad \mathcal{M}\widetilde{\mathcal{M}} = i\sqrt{2}\Lambda m, \qquad |\mathcal{M}| = |\widetilde{\mathcal{M}}|. \qquad (2.7)$$

From the dual $U(1)$ point of view, the fields $\mathcal{M}$ and $\widetilde{\mathcal{M}}$ are electrically charged and their non-zero vacuum expectation values imply electric superconductivity *à la* Higgs; the photon becomes massive and the magnetic charges suffer confinement. However, as far as the original $U(1) \subset SU(2)$ gauge theory is concerned, the superconductivity is magnetic and the confined charges are electric; this is the familiar *electric confinement* of a non-abelian gauge theory.

Likewise, at the other singularity $U = -\Lambda^2$, the effective superpotential is

$$\begin{aligned} W &= mU + (A_1(U) + A_2(U))\, \mathcal{D}\widetilde{\mathcal{D}} \\ &\approx mU + \frac{\Lambda^2 + U}{-i\sqrt{2}\Lambda} \mathcal{D}\widetilde{\mathcal{D}} \end{aligned} \qquad (2.8)$$

and the vacuum

$$U = -\Lambda^2, \qquad \mathcal{D}\widetilde{\mathcal{D}} = i\sqrt{2}\Lambda m, \qquad |\mathcal{D}| = |\widetilde{\mathcal{D}}| \qquad (2.9)$$

is a dyonic superconductor which confines particles with charges $q_e \neq q_m$; such behavior is called *oblique confinement*.

To summarize the phase structure of the Seiberg–Witten theory, the theory with $m = 0$ and unbroken $\mathcal{N} = 2$ SUSY is in the abelian Coulomb phase with a moduli space while the $\mathcal{N} = 1$ theory with $m \neq 0$ has an electric confinement phase and an oblique confinement phase, each comprising an isolated vacuum without any massless particles at all. When $m$ becomes large compared to $\Lambda$, eqs. (2.7) and (2.9) for the monopole/dyon expectation values should be renormalized to account for quantum corrections to the Kähler metric for those fields, but the two confining phases persist all the way to $m \to \infty$ limit when the $\Phi$ superfield decouples and the SW theory reduces to $\mathcal{N} = 1$ SSYM; this point is discussed in more detail in section 6.

Note that the phase structure of the $\mathcal{N} = 1$ SW theory follows from three separate low-energy effective *local* field theories covering different parts of the $\mathcal{N} = 2$ moduli space, namely the generic $U$ away from the singularities, the singular region of $U \approx +\Lambda^2$ and the other singular region of $U \approx -\Lambda^2$. The Seiberg–Witten analysis has no need for a single effective theory covering the entire moduli space — which is very fortunate since such a theory does not exist, at least not as a local field theory. Indeed, such a theory would have to include both $\mathcal{M}, \widetilde{\mathcal{M}}$ and $\mathcal{D}, \widetilde{\mathcal{D}}$ as local fields, both locally coupled to the same $U(1)$ gauge field $V_\mu$, which would require them to have purely-electric charges *in the same duality frame*, a mathematical impossibility.

Unfortunately, a domain wall between the two confining vacua — the main subject of this article — has the modulus field continuously changing its value from $U \approx +\Lambda^2$ on one side of the wall through generic values of $U$ in the middle all the way to $U \approx -\Lambda^2$ on the other side; thus one has to use three different effective theories in different regions of space to understand the whole wall. Worse, *the field configuration in the middle of the wall is not a vacuum*, so the effective theory for a generic $\mathcal{N} = 2$ SW vacuum is of questionable validity for this region, — and we do not have a better theory to replace it with.

We shall see momentarily that the answer to the last question depends on the $m/\Lambda$ ratio: For small $m \ll \Lambda$, the middle section of the BPS-saturated domain wall is basically in the adiabatically perturbed Coulomb phase of the SW theory while the monopole and the dyon fields become significant only in the two outer sections, — and no region of space has both the monopole and the dyon condensates. Consequently, using all three effective low-energy theories, we shall explicitly construct the entire field configuration of the BPS-saturated domain wall and thus confirm its existence in a fairly convincing manner. On the other hand, for large $m \geq O(\Lambda)$, the monopole and the dyon condensates spread towards each other and presumably overlap, whatever that means. Since we see no sign of a phase transition in the theory at finite $m$, the BPS-saturated domain wall presumably continues to exists all the way to $m = \infty$ *i.e.*, in the $\mathcal{N} = 1$ SSYM theory, but the detailed structure of such a wall remains out of our reach.

## 3. A Toy Model

In order to see how the small $m$ and the large $m$ regimes of the BPS domain wall differ from each other, let us now consider a simplified toy model of the Seiberg–Witten theory in which both regimes are governed by an easy to understand local $\mathcal{N} = 1$ supersymmetric effective field theory. Specifically, let us replace the electric and the magnetic charges of the SW theory with two different kinds of electric charges. That is, our model has a $U(1) \times U(1)$ gauge symmetry and the analogues of the monopole, the anti-monopole, the dyon and the anti-dyon have the following purely electric charges: $T^{(0,0)}$, $\mathcal{M}^{(+1,0)}$, $\widetilde{\mathcal{M}}^{(-1,0)}$, $\mathcal{D}^{(+1,+1)}$ and $\widetilde{\mathcal{D}}^{(-1,-1)}$; there are no magnetic charges in this model. For simplicity, we set the Fayet-Iliopoulos D-terms for the two $U(1)$ factors to zero while the superpotential is a simplified version of (2.6) combined with (2.8), namely

$$W = (T - \Lambda)\mathcal{M}\widetilde{\mathcal{M}} \;+\; (T + \Lambda)\mathcal{D}\widetilde{\mathcal{D}} \;-\; h^2 T \,. \tag{3.1}$$

where $T$ is a linearized analogue of the SW modulus $U$ and $h^2 \sim m\Lambda$.

The phase structure of our model is analogous to that of the SW theory, except for the extra massless photon. For $h = 0$ (but $\Lambda \neq 0$), the model is similar to the SW theory with the unbroken $\mathcal{N} = 2$ SUSY: There is a continuous family of degenerate vacua parametrized by the expectation value of the $T$ field. For generic values of $\langle T \rangle$, the only massless particles are the two photons and the $T$ field (and their superpartners), but for $\langle T \rangle = \pm \Lambda$, there are additional massless charged particles, namely $\mathcal{M}$ and $\widetilde{\mathcal{M}}$ or $\mathcal{D}$ and $\widetilde{\mathcal{D}}$. On the other hand, for $h \neq 0$, the model behaves like the SW theory with supersymmetry broken down to $\mathcal{N} = 1$: There are two discrete but degenerate vacua,

$$\begin{aligned}(1) \quad & T = +\Lambda, \quad \mathcal{M} = \widetilde{\mathcal{M}} = h, \quad \mathcal{D} = \widetilde{\mathcal{D}} = 0 \\ \text{and} \quad (2) \quad & T = -\Lambda, \quad \mathcal{D} = \widetilde{\mathcal{D}} = h, \quad \mathcal{M} = \widetilde{\mathcal{M}} = 0.\end{aligned} \quad (3.2)$$

Furthermore, while each vacuum leads to one massless and one massive photon, these two vacua lead to Higgsing of different linear combinations of the two $U(1)$ gauge fields; this is analogous to the two vacua of the SW theory leading respectively to *electric confinement v. oblique confinement*.

Notice that the superpotential (3.1) is single-valued and our toy model gives rise to the two vacua (3.2) without any help from some 'hidden' heavy degrees of freedom; consequently, the domain-wall solution interpolating between the two vacua can be analyzed in terms of straightforward field equations of motion. For a BPS-saturated domain wall, these equations follow from the requirement of preserving two out of four global supercharges of the $\mathcal{N} = 1$ theory, namely

$$\left(Q_\alpha - ie^{i\varphi}\sigma^3_{\alpha\dot\alpha}\overline{Q}^{\dot\alpha}\right)|\text{wall}\rangle \;=\; 0 \qquad (3.3)$$

where the $x_3$ coordinate axis is normal to the wall (*i.e.*, the wall lies in the $(x_1, x_2)$ plane) and $\varphi = \arg(\Delta W$ between the two vacua). In terms of the bosonic fields of an effective field theory and their auxiliary superpartners (generically denoted $A^i$, $V_\mu^a$ and $F^i$, $D^a$, respectively), eqs. (3.3) imply

$$\begin{aligned}V_\mu^a(x_3) &\equiv 0, \qquad D^a(x_3) \equiv 0, \\ \frac{dA^i}{dx_3} &= e^{i\varphi}F^i \equiv e^{i\varphi}g^{i\bar{j}}\frac{\partial W^*}{\partial \overline{A}^{\bar{j}}}\end{aligned} \qquad (3.4)$$

(in the axial gauge $V_3^a \equiv 0$), $g^{i\bar{j}}$ being the inverse of the quantum-corrected Kähler metric $g_{j\bar{i}}(A, A^*)$ of the scalar-field space.

For simplicity, let us take the Kähler metric of our toy model to be flat, $g_{j\bar{i}} \equiv \delta_{j\bar{i}}$ (in spite of any quantum corrections). Consequently, eqs. (3.4) for our

model become

$$\begin{aligned}
\mathcal{M} &\equiv \widetilde{\mathcal{M}}, \qquad \mathcal{D} \equiv \widetilde{\mathcal{D}}, \\
\frac{d\mathcal{M}}{dx_3} &= e^{i\varphi}\,\widetilde{\mathcal{M}}^*(T-\Lambda)^*, \\
\frac{d\mathcal{D}}{dx_3} &= e^{i\varphi}\,\widetilde{\mathcal{D}}^*(T+\Lambda)^*, \\
\frac{dT}{dx_3} &= e^{i\varphi}\left(\mathcal{M}\widetilde{\mathcal{M}}+\mathcal{D}\widetilde{\mathcal{D}}-h^2\right)^*,
\end{aligned} \quad (3.5)$$

where $\varphi = \arg(\Delta W = 2\Lambda h^2)$, assuming the domain wall solution asymptotes to the first vacuum of (3.2) for $x \to -\infty$ and to the second vacuum for $x \to +\infty$.

Eqs. (3.5) and (3.2) are invariant under global phase changes

$$\mathcal{M} \to e^{i\beta}\mathcal{M}, \quad \widetilde{\mathcal{M}} \to e^{i\beta}\widetilde{\mathcal{M}}, \quad \mathcal{D} \to e^{i\beta}\mathcal{D}, \quad \widetilde{\mathcal{D}} \to e^{i\beta}\widetilde{\mathcal{D}}, \quad h \to e^{i\beta}h,$$
$$T \to e^{i\gamma}T, \quad \Lambda \to e^{i\gamma}\Lambda,$$

so without loss of generality we take the parameters $h$ and $\Lambda$ to be real and positive. Another important symmetry is the charge conjugation, which (for real $h$ and $\Lambda$) acts on all the scalar fields of the theory by complex conjugation. Since a BPS-saturated domain wall between two given vacua should be unique (if it exists at all), we should look for a solution in which all fields are *real* functions of the $x_3$ coordinate.

Solving eqs. (3.5) for the real fields is a fairly straightforward procedure. First, the equations for the $\mathcal{M}$ and $\mathcal{D}$ fields imply

$$\frac{d}{dx_3}\log(\mathcal{M}\mathcal{D}) = 2T, \qquad \frac{d}{dx_3}\log\frac{\mathcal{D}}{\mathcal{M}} = 2\Lambda, \qquad (3.6)$$

which allows us to express the entire field configuration in terms of just one real function $\xi(x_3)$ according to

$$\begin{aligned}
\mathcal{M}(x_3) = \widetilde{\mathcal{M}}(x_3) &= h\exp\left(-\tfrac{1}{2}\xi(x_3)-\Lambda x_3\right), \\
\mathcal{D}(x_3) = \widetilde{\mathcal{D}}(x_3) &= h\exp\left(-\tfrac{1}{2}\xi(x_3)+\Lambda x_3\right), \\
T(x_3) &= -\frac{1}{2}\frac{d\xi}{dx_3}.
\end{aligned} \quad (3.7)$$

(Without loss of generality, we have placed the origin of the $x_3$ coordinate in the middle of the wall.) In terms of the $\xi(x_3)$, eq. (3.5) for the $T$ field becomes

$$\frac{d^2\xi}{dx_3^2} = -2\frac{dT}{dx_3} = 2\left(h^2 - \mathcal{M}^2 - \mathcal{D}^2\right) \qquad (3.8)$$
$$= 2h^2\left(1 - 2\cosh(2\Lambda x_3)\exp(-\xi)\right),$$

while the boundary conditions

$$\xi(x_3 \to \pm\infty) = 2\Lambda|x_3| \qquad (3.9)$$

assure the field configuration asymptotically approaches the vacua (3.2) as one goes away from the domain wall.

From the general form of eq. (3.8), it is clear that there is a unique solution $\xi(x_3)$ for any $h, \Lambda > 0$, although for generic values of $h$ and $\Lambda$, the solution is not expressible in closed form in terms of familiar functions. However, for the special case of $h^2 = 2\Lambda^2$, the solution simplifies to

$$\xi(x_3) = 2\log\left(2\cosh\Lambda x_3\right) \qquad (3.10)$$

and hence

$$\mathcal{M} = \widetilde{\mathcal{M}} = \frac{h}{1 + e^{2\Lambda x_3}},$$
$$\mathcal{D} = \widetilde{\mathcal{D}} = \frac{h}{1 + e^{-2\Lambda x_3}}, \qquad (3.11)$$
$$T = -\Lambda\tanh\Lambda x_3.$$

Similarly, in the limit $h \gg \Lambda$, the solution becomes

$$\xi(x_3) \approx \log\left(2\cosh 2\Lambda x_3\right) \qquad (3.12)$$

and hence

$$\mathcal{M} = \widetilde{\mathcal{M}} \approx \frac{h}{\sqrt{1 + e^{4\Lambda x_3}}},$$
$$\mathcal{D} = \widetilde{\mathcal{D}} \approx \frac{h}{\sqrt{1 + e^{-4\Lambda x_3}}}, \qquad (3.13)$$
$$T \approx -\Lambda\tanh 2\Lambda x_3.$$

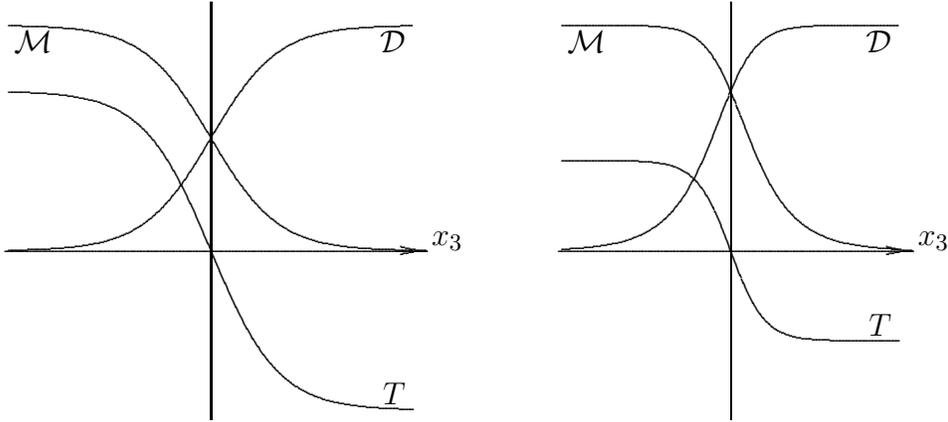

Figure 1: *Field profiles for the domain wall solution of the toy model. The profiles on the left picture correspond to $h^2 = 2\Lambda^2$; on the right picture, $h \gg \Lambda$. Note the similarity of the two pictures.*

The two solutions are plotted in Figure 1, which shows their qualitative similarity to each other; both walls have similar distance scales $O(1/\Lambda)$ that govern the $x_3$ dependence of all the fields and there are no interesting features at either longer or shorter distances. Naturally, we expect the same behavior from the domain wall for any $h \gtrsim O(\Lambda)$.

On the other hand, the $h \ll \Lambda$ regime of the domain wall solution is very different. Indeed, assuming a rather mild inequality

$$\exp\left(\frac{\Lambda^2}{h^2}\right) \gg 1, \tag{3.14}$$

we approximate eq. (3.8) as${}^\star$

$$\frac{d^2\xi}{dx_3^2} \approx 2h^2\left(1 - \exp(2\Lambda|x_3| - \xi)\right), \tag{3.15}$$

---

${}^\star$ This approximation is based upon the lower bound $\xi > (\Lambda/h)^2$ (which follows from $\frac{d^2\xi}{dx_3^2} < 2h^2$ while $\xi \geq 2\Lambda|x_3|$). Consequently, the assumption (3.14) implies $e^\xi \gg 1$ and hence $1 - 2\cosh(2\Lambda x_3)\exp(-\xi) = 1 - \exp(2\Lambda|x_3| - \xi) - \exp(-2\Lambda|x_3| - \xi) \approx 1 - \exp(2\Lambda|x_3| - \xi)$.

which can be solved in terms of the integral equation

$$2h|x_3| = \int_{\xi(x_3)-2\Lambda|x_3|}^{\sigma_0} \frac{d\sigma}{\sqrt{\sigma + e^{-\sigma} - 1}}, \qquad (3.16)$$

where $\sigma_0 + e^{-\sigma_0} - 1 = (\Lambda/h)^2$; in light of the assumption (3.14), $\sigma_0 \approx (\Lambda/h)^2 + 1$. The integral in eq. (3.16) can be evaluated explicitly whenever its lower limit $\xi(x_3) - 2\Lambda|x_3|$ is either large or small, which leads to

$$\xi(x_3) = \frac{\Lambda^2}{h^2} + h^2 x_3^2 + 1 + O\left(\exp\left[-\left(\tfrac{\Lambda}{h} - h|x_3|\right)^2\right]\right) \qquad (3.17)$$

for $|x_3| < \Lambda h^{-2} - O(h^{-1})$ and

$$\begin{aligned}
\xi(x_3) &= 2\Lambda|x_3| + 2C\exp\left[-\sqrt{2}h\left(|x_3| - \Lambda h^{-2}\right)\right] \\
&\quad + O\left(\exp\left[-2\sqrt{2}h\left(|x_3| - \Lambda h^{-2}\right)\right]\right)
\end{aligned} \qquad (3.18)$$

($C \approx 0.3312$ is a numerical constant) for $|x_3| > \Lambda h^{-2} + O(h^{-1})$. In the intervening ranges of $x_3$ near $\pm\Lambda h^{-2}$,

$$\xi(x_3) \approx 2\Lambda|x_3| + O(1) \quad \text{for } \left||x_3| - \Lambda h^{-2}\right| \lesssim O(h^{-1}) \qquad (3.19)$$

but a numerical evaluation of the integral (3.16) is required for a more accurate answer. Figure 2 shows the field profiles resulting from such a numerical integration.

Altogether, the domain wall of the model with $h \ll \Lambda$ has a five-layer sandwich-like configuration:

1. The Left Asymptotic region at $x_3 < -\Lambda h^{-2} - O(h^{-1})$ where

$$\begin{aligned}
\mathcal{M} = \widetilde{\mathcal{M}} &\approx h - Ch\exp\left(-\sqrt{2}h(|x_3| - \Lambda h^{-2})\right), \\
\mathcal{D} = \widetilde{\mathcal{D}} &\approx h\exp\left(-2\Lambda|x_3|\right) \approx 0 \\
\text{and} \quad T &\approx \Lambda - \sqrt{2}Ch\exp\left(-\sqrt{2}h(|x_3| - \Lambda h^{-2})\right).
\end{aligned} \qquad (3.20)$$

This is the exponential tail of the domain wall into the domain of the electric confinement phase[†] at $x_3 \to -\infty$.

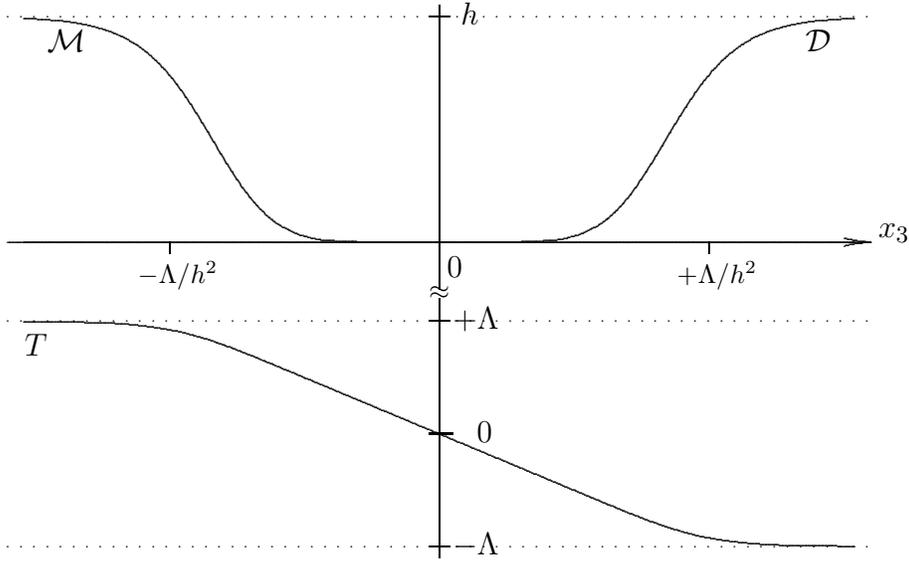

Figure 2: *The domain wall profile for the toy model with $\Lambda = 5h$ shows the five-layer structure of the $\Lambda \gg h$ limit.*

2. The Left Transitional region at $-O(h^{-1}) < x_3 + \Lambda h^{-2} < +O(h^{-1})$ where

$$T \approx \Lambda - O(h), \qquad \mathcal{D} = \widetilde{\mathcal{D}} = O(e^{-\Lambda^2/h^2}) \approx 0, \qquad \mathcal{M} = \widetilde{\mathcal{M}} = O(h)$$
(3.21)

but the exact profiles of the $\mathcal{M}(x_3)$ and $\Lambda - T(x_3)$ are rather complicated. Qualitatively, the monopole fields $\mathcal{M}$ and $\widetilde{\mathcal{M}}$ 'switch off' *i.e.*, decrease from $h$ to almost zero while the $T$ field does not deviate far from its vacuum value $+\Lambda$ (note $\Lambda \gg h$); the dyon fields $\mathcal{D}$ and $\widetilde{\mathcal{D}}$ remain negligible throughout this region.

---

† We name the phases of our toy model after their SW analogues. Likewise, we call $\mathcal{M}$ the 'monopole' field, $\mathcal{D}$ the 'dyon', *etc.*, *etc.*

3. The Middle Layer at $-\Lambda h^{-2} + O(h^{-1}) < x_3 < +\Lambda h^{-2} - O(h^{-1})$ where

$$\begin{aligned} \mathcal{M} &= \widetilde{\mathcal{M}} \approx h \exp\left(-h^2 \left(x_3 + \Lambda h^{-2}\right)^2\right) \approx 0, \\ \mathcal{D} &= \widetilde{\mathcal{D}} \approx h \exp\left(-h^2 \left(x_3 - \Lambda h^{-2}\right)^2\right) \approx 0 \\ \text{and} \quad T &\approx -h^2 \, x_3\,. \end{aligned} \quad (3.22)$$

In this layer, both the monopole and the dyon fields are negligible while the modulus field $T$ *slowly* interpolates between its two vacuum values $+\Lambda$ and $-\Lambda$.

4. The Right Transitional region at $-O(h^{-1}) < x_3 - \Lambda h^{-2} < +O(h^{-1})$ where the 'monopole' fields are negligible, the modulus $T$ is close to its expectation value $-\Lambda$ in the domain of the oblique confinement phase and the dyon fields $\mathcal{D}$ and $\widetilde{\mathcal{D}}$ 'switch on' *i.e.*, increase from almost zero to almost $h$,

$$\mathcal{M} = \widetilde{\mathcal{M}} = O(e^{-\Lambda^2/h^2}) \approx 0, \qquad T \approx -\Lambda + O(h), \qquad \mathcal{D} = \widetilde{\mathcal{D}} = O(h). \quad (3.23)$$

5. The Right Asymptotic region at $x_3 > +\Lambda h^{-2} + O(h^{-1})$ or the exponential tail of the domain wall into the oblique confinement domain at $x_3 = +\infty$.

Notice that for $h \ll \Lambda$, the characteristic length scale $O(1/h)$ of significant field change in the two transition regions (2 and 4 above) is much shorter than the thickness $2\Lambda/h^2$ of the middle layer (3) of the domain wall. Indeed, in the $h \to 0$ limit of the wall, the middle layer becomes so thick that it acts almost like a separate phase of the theory, greatly resembling the 'Coulomb' phase of the $h = 0$ model where $\langle \mathcal{M} \rangle = \langle \widetilde{\mathcal{M}} \rangle = \langle \mathcal{D} \rangle = \langle \widetilde{\mathcal{D}} \rangle = 0$ while the modulus field $T$ has an arbitrary vacuum expectation value. *In bulk*, this Coulomb phase becomes energetically unstable — as well as non-supersymmetric — in presence of a non-zero O'Raighfeartaigh F-term $h^2 T$, however small, but *in a layer of a finite thickness*, a small O'Raighfeartaigh term is just an adiabatic perturbation that turns on a gradient $\nabla T = -h^2$ of the modulus field. The perturbed Coulomb phase remains stable and invariant under two out of four supercharges — and clearly this is precisely the field configuration in the middle layer (3.22) of the $h \ll \Lambda$ domain wall.

Altogether, our field configuration is best thought of not as two phases separated by a multi-layer domain wall but rather as a sequence of *three domains of distinct phases* separated by two transitional regions. In SW-inspired terminology, the three phases are respectively the electric confinement phase, the perturbed Coulomb phase and the oblique confinement phase; the two confining phases extend to $x_3 = \mp\infty$ while the perturbed Coulomb phase has a finite thickness $2\Lambda h^{-2}$. The two transition regions between either confining phase and the (perturbed) Coulomb phase are in a sense domain walls on their own rights, well separated from each other and from the other confining phase and much thinner than the Coulomb phase domain between them.

Clearly, this three-phase structure of a BPS-saturated domain wall is by no means a peculiar feature of our toy model. Instead, it is generic to all theories where a continuous family of exactly degenerate vacua collapses to several isolated vacua when one adds *small* F-terms $W_F$ for the moduli fields of the theory (*e.g.*, $W_F = h^2 T$): As long as such F-terms are small, the adiabatic perturbation of the generic-moduli phase (*e.g.*, the Coulomb phase of the SW theory or our toy model) is stable and invariant under two supercharges in a layer of a finite thickness in which the moduli fields have gradients $\nabla T^i = e^{i\varphi} g^{i\bar{j}} (\partial W_F / \partial T^j)^*$ (*cf.* eq. (3.4)). Within the domain of this phase, one may ignore the fields that become light only at some special points or subspaces of the moduli space.

In all such theories, a BPS domain wall between two bulk-stable phases has a three-domain structure with the (perturbed) generic-moduli phase occupying the middle domain. The relatively thin (for small O'Raighfeartaigh terms) transition regions between this domain and the bulk-stable phases on either side of the domain wall are characterized by moduli fields having values within $O(h)$ of the appropriate bulk-stable phase's VEVs and also by non-trivial values of the light fields and condensates present in that particular phase (*e.g., either* $\mathcal{M}$ and $\widetilde{\mathcal{M}}$ or $\mathcal{D}$ and $\widetilde{\mathcal{D}}$). However, one may safely ignore the fields that become light in any other phase of the theory; for example, in the toy model, the $\mathcal{D}$ and $\widetilde{\mathcal{D}}$ fields play no role in the transition between the Coulomb and the electric-confinement phases while the $\mathcal{M}$ and $\widetilde{\mathcal{M}}$ fields are as good as absent from the Coulomb-to-the-oblique-confinement transition region.

The above picture describes the $h \ll \Lambda$ regime of *small* O'Raighfeartaigh terms, but increasing $h$ gives rise to two major complications: First, perturbing the generic-moduli phase too strongly may change its nature; indeed, the $O(h^2)$ gradients of the moduli amount to a momentum, so all particles with masses up to $O(h^2/\Lambda)$ should be included in the effective theory. Second, the $O(\Lambda/h^2)$ thickness of the middle domain shrinks with $h$ at a faster rate than the $O(1/h)$ characteristic length scale of the transition regions. Eventually, for $h = O(\Lambda)$, the middle domain disappears under the (inner) tail ends of the two transition regions and the latter merge into a single featureless wall where all fields are present more or less throughout the wall's thickness. For example, in our toy model, both 'monopole' and 'dyon' fields have $O(h)$ values in the middle of the domain wall, *cf.* figure 1. Generically, in a more complicated theory with a similar three-domain structure of a BPS domain wall in the $h \ll \Lambda$ regime, we expect the $h \geq O(\Lambda)$ regime of the wall to be quite different.

## 4. Seiberg–Witten Domain Walls in the $\mathcal{N} = 2$ limit $m \ll \Lambda$

Applying the general conclusions of the previous subsection to the Seiberg–Witten theory[13] immediately tells us that for *small m*, the SW domain wall has a three-domain structure with the middle domain occupied by the Coulomb phase adiabatically perturbed by the O'Raighfeartaigh term $mU$.[26] Since $U$ is the only light scalar field in this phase, the BPS equations (3.4) reduce to a single equation

$$\frac{dU}{dx_3} = m^* e^{i\varphi} g^{U\overline{U}}. \tag{4.1}$$

Without loss of generality, we assume real positive $m$ and $\Lambda$, which makes for $e^{i\varphi} = -1$ and real $U(x_3)$ throughout the Coulomb domain (and indeed the whole domain wall, as we shall see in a moment). Eq. (4.1) has an obvious formal solution

$$m x_3 = \int_U^0 dU' \, g_{U\overline{U}}(U'), \qquad (4.2)$$

in terms of the Kähler metric (2.1). The latter follows from the elliptic curve's periods (2.2), which in turn can be expressed in terms of hypergeometric functions.[27] Unfortunately, the resulting formula is too complicated to evaluate the integral (4.2) analytically. Figure 3a shows the numerical behavior of the metric for real $U$ between $-\Lambda^2$ and $+\Lambda^2$: It looks rather flat, except for mild logarithmic singularities at the two ends of the graph; analytically

$$g_{U\overline{U}}(U) \approx \frac{1}{16\pi^2 \Lambda^2} \log \frac{64 \Lambda^4}{\Lambda^4 - U^2} \qquad (4.3)$$

for $U \to \pm\Lambda^2$. Naively, this logarithmic behavior seems to be limited to $U$ being fairly close to $\pm\Lambda^2$, but actually, the approximation (4.3) is amazingly accurate throughout the $-\Lambda^2$ to $+\Lambda^2$ range: The error is barely 5% in the middle of the range (*i.e.*, $U \approx 0$) and decreases towards its ends. At the moment, we have no explanation for this unexpected accuracy but only a conjecture that it reflects some important feature of the dual instanton expansion of the SW theory; this is a subject for future research.

Substituting the approximate metric (4.3) into eq. (4.2) gives us the (approximate) field profile of the Coulomb domain within the BPS domain wall:

$$16\pi^2 m \, x_3 = \log \frac{\Lambda^2 + U}{\Lambda^2 - U} - \frac{U}{\Lambda^2} \log \frac{64 e^2 \Lambda^4}{\Lambda^4 - U^2}; \qquad (4.4)$$

as expected, the Coulomb domain has finite thickness $w = \frac{1 + \log 4}{4\pi^2 m} \approx 0.0605 \, m^{-1}$. For a better accuracy, we have numerically integrated eq. (4.2) using the exact metric (2.1): The resulting field profile is presented on figure 3b and the Coulomb domain's thickness turns out to be $w = 0.0625\, m^{-1}$.

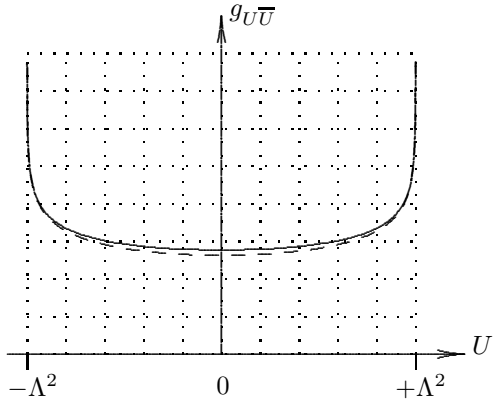
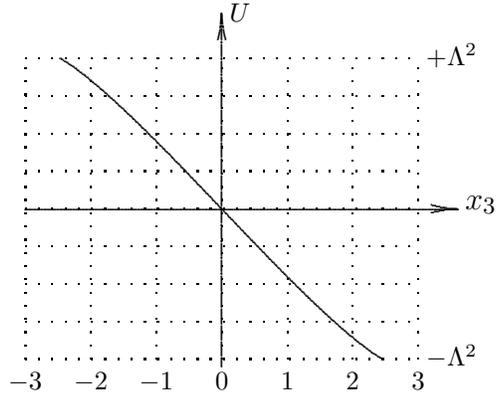

Figure 3a: *Metric for the Seiberg–Witten modulus $U$. The solid line plots the actual metric (2.1), the dashed line — the approximate metric (4.3); the metric $g_{U\overline{U}}$ is in units of $0.01\Lambda^{-2}$ per grid line.*

Figure 3b: *Field profile $U(x_3)$ for the perturbed Coulomb domain inside the Seiberg–Witten domain wall (for $m \ll \Lambda$); the $x_3$ is in units of $1/8\pi^2 m$.*

Next, consider the transition regions between the Coulomb and the confining domains. According to general rules, the monopole fields $\mathcal{M}$ and $\widetilde{\mathcal{M}}$ which play a key role in electric-confinement phase are also important in the transition region between that phase's domain and the Coulomb domain, but they do not show up in the other regions of the wall. Likewise, the dyon fields $\mathcal{D}$ and $\widetilde{\mathcal{D}}$ are limited to the oblique-confinement phase and the transition between that phase and the Coulomb phase. Hence, *the monopole and the dyon fields are spatially segregated from each other* and thus can be taken care of by separate but local effective field theories.

Because of the obvious symmetry between the two transition regions, we shall limit our discussion to the electric-confinement-to-Coulomb transition on the left side of the wall. The effective field theory governing this transition region has $T$, $\mathcal{M}$ and $\widetilde{\mathcal{M}}$ fields and the superpotential (2.6); consequently, the BPS equations for this region are

$$\begin{aligned}
g_{U\overline{U}} \frac{dU}{dx_3} &= -m + \frac{i}{\sqrt{2}\Lambda} \mathcal{M}^* \widetilde{\mathcal{M}}^*, \\
g_{\mathcal{M}} \frac{d\mathcal{M}}{dx_3} &= \frac{\Lambda^2 - U^*}{i\sqrt{2}\Lambda} \widetilde{\mathcal{M}}^*, \\
g_{\mathcal{M}} \frac{d\widetilde{\mathcal{M}}}{dx_3} &= \frac{\Lambda^2 - U^*}{i\sqrt{2}\Lambda} \mathcal{M}^*, \\
|\mathcal{M}| &\equiv |\widetilde{\mathcal{M}}|,
\end{aligned} \quad (4.5)$$

where $g_{\mathcal{M}}$ is a short-hand notation for the Kähler metric $g_{\mathcal{M}\overline{\mathcal{M}}} = g_{\widetilde{\mathcal{M}}\overline{\widetilde{\mathcal{M}}}}$ for the monopole fields. In the $\mathcal{N} = 2$ limit of the SW theory, this metric is $U$-independent (hypermultiplets' metric does not depend on vector-multiplet moduli) and can be taken to be canonical, $g_{\mathcal{M}} = 1$.

The boundary conditions for eqs. (4.5) at $x_3 = -\infty$ are given by the vacuum expectation values (2.7). Note that both the boundary conditions and the equations (4.5) themselves are invariant under complex conjugation of the fields $U$, $\mathcal{M}$ and $i\widetilde{\mathcal{M}}$, which means that the unique BPS-saturated domain wall solution should involve real fields $U$ and $\mathcal{M}$ and imaginary $\widetilde{\mathcal{M}} \equiv i\mathcal{M}$ (modulo a gauge symmetry of the $\mathcal{M}$ and $\widetilde{\mathcal{M}}$ fields). Let us therefore substitute

$$-i\widetilde{\mathcal{M}} = \mathcal{M} = h e^{-\alpha/2}, \qquad U = \Lambda^2 - \sqrt{\tfrac{1}{2}} \Lambda \beta \quad (4.6)$$

where $h^2 = \sqrt{2}\Lambda m$. In terms of our new variables $\alpha(x_3)$ and $\beta(x_3)$, eqs. (4.5) become

$$\begin{aligned}
\frac{d\alpha}{dx_3} &= \beta, \\
\frac{\Lambda g_{U\overline{U}}(\beta)}{\sqrt{2}} \frac{d\beta}{dx_3} &= m\left(1 - e^{-\alpha}\right).
\end{aligned} \quad (4.7)$$

Weirdly, this equation system has the form of a classical Hamiltonian or Lagrangian system where $x_3$ plays the role of time, $\alpha$ is the canonical position variable, $\beta$ is the velocity,

$$\rho = \int \frac{\Lambda}{\sqrt{2}} g_{U\bar{U}}(\beta) \, d\beta \approx \frac{\beta}{16\sqrt{2}\pi^2\Lambda} \log \frac{32\sqrt{2}e\Lambda}{\beta} \qquad (4.8)$$

is the canonical momentum conjugate to $\alpha$,[★]

$$K = \int \beta d\rho \approx \frac{\beta^2}{64\sqrt{2}\pi^2\Lambda} \log \frac{2048e\Lambda^2}{\beta^2} \qquad (4.9)$$

is the kinetic energy and

$$V = -m\left(\alpha - 1 + e^{-\alpha}\right) \qquad (4.10)$$

is the potential energy. The physical meaning of this Hamiltonian system is at the moment completely obscure, but it does provide us with a convenient language for describing the transition: The domain of the electric confinement phase lies in the asymptotic past ($x_3 \to -\infty$) where $\alpha = 0$, $\beta = 0$ and both the potential and the kinetic energy vanish. Note that $\alpha = 0$ is at the maximum of the potential (4.10), so as the time $x_3$ goes on, the system slides off this 'potential hill' and both the coordinate $\alpha$ and the velocity $\beta$ begin to grow according to the law of energy conservation

$$K(\beta) + V(\alpha) = \text{const} = 0. \qquad (4.11)$$

Eventually, as $\alpha$ grows large, the 'force' $F = -dV/d\alpha$ becomes more or less constant, $F \approx m$, and the mechanical system ends up in free fall, $\rho \approx mx_3 + \text{const}$.

---

[★] The approximation in this formula corresponds to $g_{U\bar{U}} \approx (4\pi\Lambda)^{-2} \log(32\sqrt{2}\Lambda/\beta)$, cf. eq. (4.3).

From the domain wall's point of view, the free fall in the asymptotic future describes the domain of the perturbed Coulomb phase. Indeed, the free-fall equation $\rho \approx mx_3 + \text{const}$ means

$$16\pi^2 m\, x_3 \;=\; \text{const} \;+\; \frac{\Lambda^2 - U}{\Lambda^2}\, \log \frac{32 e \Lambda^2}{\Lambda^2 - U}, \tag{4.12}$$

which agrees with the $\Lambda^2 - U \ll \Lambda^2$ limit of the Coulomb-domain formula (4.4). At the same time, energy conservation (4.11) implies

$$\alpha \;\approx\; 1 + \tfrac{1}{m} K(\beta) \;=\; 1 \;+\; \left(\frac{\Lambda^2 - U}{4\pi \Lambda h}\right)^2 \left(\log \frac{32\Lambda^2}{\Lambda^2 - U} + \frac{1}{2}\right), \tag{4.13}$$

which becomes large for $U \le \Lambda^2 - O(4\pi\Lambda h)$. Consequently, the monopole condensate $\mathcal{M} = h e^{-\alpha/2}$ peters out, exactly as it should do as one goes into the Coulomb phase domain.

The exact solution to the equations of motion (4.7) is given by the integral

$$x_3 \;=\; \int \frac{d\alpha}{\beta} \tag{4.14}$$

where $\alpha$ and $\beta$ are related to each other via eq. (4.11). Similar to the toy model, we have analytic formulæ for this integral in the asymptotic regimes of either small or large $\alpha$ but a numerical evaluation is required for the intermediate range; the field profiles resulting from such a numerical integration are presented on figure 4. Analytically, eqs. (4.12) and (4.13) describe the asymptotic form of the field profiles on the Coulomb-side tail end of the transition, $x > -\tfrac{1}{2}w + O(1/8\pi h)$ ($w \approx 0.0625 m^{-1}$ being the domain wall's thickness in the $\Lambda \gg m$ limit); at the confinement-phase tail end, $x < -\tfrac{1}{2}w - O(1/8\pi h)$ the fields behave according to

$$\begin{aligned}
U &\approx \Lambda^2 \;-\; (32\Lambda^2)\, \exp\left(-\tfrac{1}{2} - (6\pi h(x_0 - x))^{2/3}\right), \\
\mathcal{M} &\approx h \;-\; (\tfrac{4\sqrt{2}}{\pi}\Lambda)\, (6\pi h(x_0 - x))^{1/3}\, \exp\left(-\tfrac{1}{2} - (6\pi h(x_0 - x))^{2/3}\right),
\end{aligned} \tag{4.15}$$

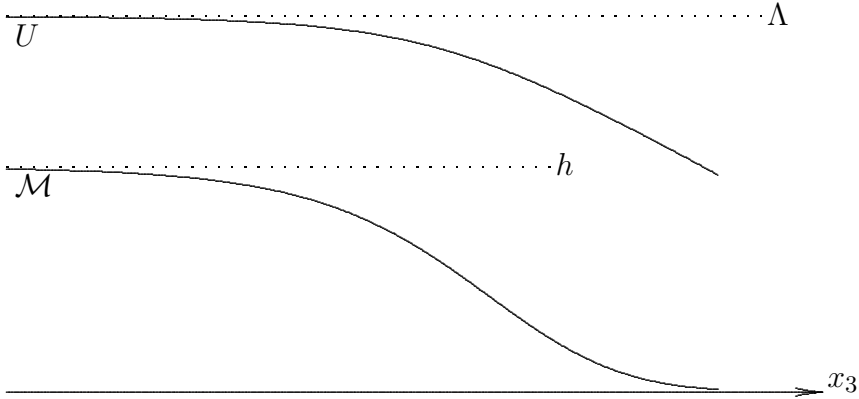

Fugure 4: *Field profiles for the transitional region of the Seiberg–Witten domain wall ($m = \Lambda/2000$). The electric-confinement domain is to the left of the plot, the Coulomb domain is to the right.*

for some constant $x_0 \approx -\frac{1}{2}w$.[†]

We conclude this section with an overview of the entire Seiberg–Witten domain wall put together. Figure 5 shows the profiles of the $U$, $\mathcal{M}$ and $\mathcal{D}$ fields for $m = \Lambda/2000$. Notice that the transition regions are unexpectedly thick for such a small $m/\Lambda$ ratio. For smaller masses, the transition regions become thicker in absolute terms — their thickness is $O(1/h)$ and thus increases as $1/\sqrt{m}$ — but the Coulomb phase domain becomes thicker at the faster rate $w \propto 1/m$,

---

† More precisely,

$$U \approx \Lambda^2 - (32\Lambda^2) \exp\left(\tfrac{1}{2} - t^2\right)\left(1 + O(e^{-t^2})\right),$$
$$\mathcal{M} \approx h - (\tfrac{4\sqrt{2}}{\pi}\Lambda)\, t \exp\left(\tfrac{1}{2} - t^2\right)\left(1 + O(e^{-t^2})\right), \qquad (4.15)'$$
$$x \approx -\frac{w}{2} + \frac{1}{6\pi h}\left(-(t^3 - \tfrac{3}{2}t) + (t_0^3 - \tfrac{3}{2}t_0) + C_1 + O(e^{-t^2})\right),$$

where $t$ is an auxiliary parameter, $t_0$ is the larger of the two solutions of the transcendental equation

$$t_0 e^{-t_0^2} = C_2 \frac{h}{\Lambda}$$

and $C_1 \approx 0.31$ and $C_2 \approx 0.115$ are numerical constants.

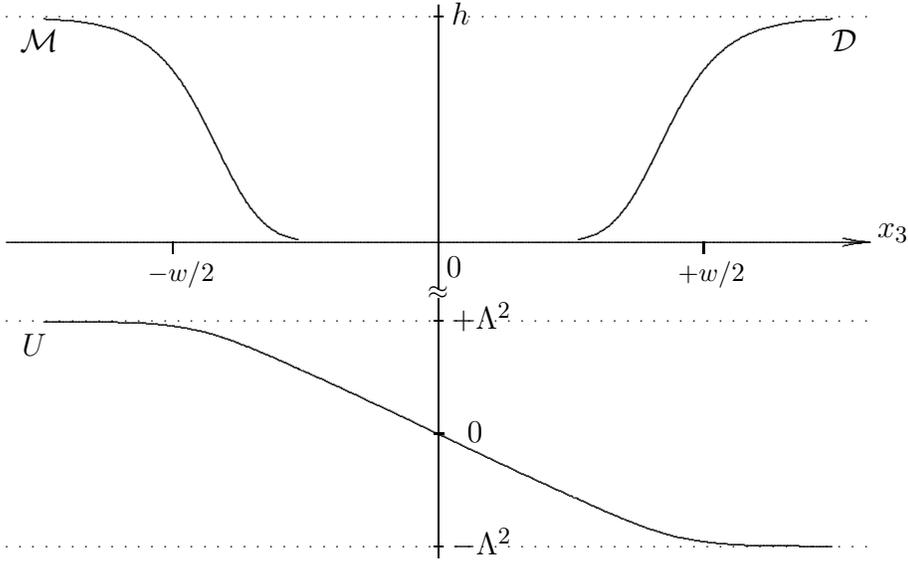

Figure 5: *Field profiles for the whole Seiberg–Witten domain wall ($m = \Lambda/2000$).*

so in relative terms, the transition regions become thinner. For example, for $m = \Lambda/2,000,000$, 96% of the total wall's thickness is taken by the Coulomb-phase domain while each transition region takes only 2%.[‡] On the other hand, for larger $m/\Lambda$ ratios, the transitional regions encroach even more on the Coulomb phase domain until they take over the whole domain wall for $m \sim \Lambda/400$. Indeed, in order to have $\alpha \gtrsim 9$ (*i.e.*, $\mathcal{M} \lesssim h/100$) while at the same time $\Lambda^2 - U \lesssim \Lambda^2$, eq. (4.13)[§] requires $\Lambda \gtrsim 17h$ *i.e.*,

$$\textbf{small } m \textbf{ means} \quad \frac{m}{\Lambda} \lesssim \frac{1}{400}. \tag{4.16}$$

As of this writing, this surprisingly stringent requirement for the validity of the

---

‡ For the purpose of this comparison, we define the transition regions as regions in which the monopole or dyon fields change from $0.99\,h$ on the confinement-phase side to $0.01\,h$ on the Coulomb-phase side.

§ Strictly speaking, eq. (4.13) is only accurate for $\Lambda^2 - U \ll \Lambda^2$, but we estimate the error in extrapolating this formula to $U \sim 0$ as being no worse than about 20%.

$m \to 0$ picture of the SW domain wall remains a puzzle.¶

## 5. Limitations of the Seiberg–Witten Theory

Beyond the very-small-mass limit (4.16), the two transitional regions of the SW domain wall not only take over the entire wall's thickness but also overlap each other. Consequently, *the monopole and the dyon condensates now coexist in the same region of space.* In the toy model with $h \gtrsim \Lambda$, similar coexistence of the $\langle \mathcal{M} \rangle$ and $\langle \mathcal{D} \rangle$ condensates was a mere calculational inconvenience since both the 'monopoles' and the 'dyons' were actually electrically charged fields of the same local field theory. In the Seiberg–Witten case however, we have a real problem: The monopoles and the dyons are now true to their names and thus cannot be simultaneously described as local fields; consequently, *we simply do not have a theory to describe coexisting condensates $\langle \mathcal{M} \rangle$ and $\langle \mathcal{D} \rangle$.*

Naturally, before we bust our brains trying to develop such a theory, we would like to know whether it would actually solve the problem at hand. Specifically, the question is whether an effective theory somehow comprising five chiral multiplets $\mathcal{M}$, $\widetilde{\mathcal{M}}$, $\mathcal{D}$, $\widetilde{\mathcal{D}}$ and $U$ would be able to adequately describe the SW domain wall for $m \gtrsim O(\Lambda)$? The answer to this question turns out to be negative: The overlapping monopole and dyon condensates are only a part of our problem as the entire Seiberg–Witten theory reaches the limit of its validity.

---

¶ One reason for the smallness of this limitation is the unphysical normalization of the holomorphic mass parameter $m$. In QCD-like terms, the $\Phi$ superfield describes an $SU(2)$ triplet of fermions and scalars whose canonically normalized running current mass is

$$M_c(p^2) \;=\; \frac{m}{g_{\Phi\overline{\Phi}}(p^2)} \;=\; m g_{SU(2)}^2(p^2) \;=\; \frac{4\pi^2 m}{\log(p^2/\Lambda^2)} \,.$$

In QCD, the running stops at $p^2 \sim (2\pi\Lambda)^2$; assuming similar pattern in the Seiberg–Witten theory, we have $M_c \sim 10m$.

Naively, this normalization issue suggest the onset of the small-$m$ regime of the SW theory for $m \lesssim \Lambda/10$. Instead, the small-$m$ picture of the SW domain wall does not emerge until the mass $m$ becomes 40 times smaller than that; we do not know why.

The basic limitation of the Seiberg–Witten theory is that it's basically the effective low-energy — *i.e.*, long-distance — theory of the massless fields, namely the $U$ modulus and its $\mathcal{N} = 2$ superpartners. It gives us the exact two-derivative interactions of those fields (encoded in the $U$ dependent gauge coupling (2.3) and metric (2.1)), but it tells us nothing about the four-derivative, six-derivative, *etc.*, interactions because they are not protected by the $\mathcal{N} = 2$ supersymmetry. Such higher-derivative interactions are literally irrelevant to the long-distance limit of the theory, but they are very much relevant to the finite-size field configurations such as finite-thickness domain walls.

Indeed, consider a BPS-saturated domain wall in a generic $\mathcal{N} = 1$ effective field theory. The exact BPS equations

$$V_a^\mu(x_3) \;=\; 0, \qquad D_a(x_3) \;=\; 0, \qquad \frac{dA^i}{dx_3} \;=\; e^{i\varphi}\, F^i \qquad (5.1)$$

follow from eq. (3.3), but they are of little practical use without specific formulæ for the auxiliary fields $D_a$ and $F^i$ in terms of the physical scalar fields $A^i$. Such formulæ follow from the generating functional $\Gamma$ (also known as 'the effective classical Lagrangian') of the effective theory. Truncating $\Gamma$ to terms with four or less super-derivatives (*i.e.*, two derivatives or four fermions or two $F$'s) results in $F^i = g^{i\bar{j}} W_{\bar{j}}^*$ and hence eqs. (3.4) for the domain wall, but allowing for higher-derivative terms would make for much more complicated formulæ. In particular, at the six-super-derivative level one has quadratic equations for the auxiliary fields $F^i$ while at still higher derivative level the $F^i$ become propagating fields governed by differential (rather than merely polynomial) equations. Examples of such equations are presented in Appendix A but their specific form is not germane to the present discussion; it suffices to say that they are quite complicated. Consequently, the exact equations for the BPS domain walls' profiles depend in a complicated way on the higher-derivative terms in the effective Lagrangian $\Gamma$. Generally, one needs to know all the terms in $\Gamma$ that are important at the distance scale of the wall's thickness: For the thick walls, only the superpotential, the Kähler function and the gauge couplings are important and the wall's profile can be described by the approximate eqs. (3.4); the thin walls are much more complicated.

The Seiberg–Witten theory is the infrared limit of the $\mathcal{N} = 2$ SSYM theory; it is valid at distances $\ell > 1/\Lambda$ but breaks down at shorter distances. Indeed, the SSYM theory has various "hadrons" with masses $O(\Lambda)$ — not just the monopoles $\mathcal{M}, \widetilde{\mathcal{M}}$ and the dyons $\mathcal{D}, \widetilde{\mathcal{D}}$, but many other BPS saturated and unsaturated particles as well. Once all those hadrons are integrated out of the effective low-energy theory, the resulting effective SW Lagrangian $\Gamma$ acquires all kinds of higher-derivative terms, which are irrelevant in the very long distance limit but become very important at $\ell \lesssim O(1/\Lambda)$. Consequently, we may use the long-distance SW theory to describe domain walls that happen to be much thicker than $O(1/\Lambda)$ — and the thicker the wall, the more accurate the description — but the thinner walls are governed by the shorter-distance theory and cannot be adequately described in the SW terms. The wall's thickness (in $1/\Lambda$ units) decreases with the $\mathcal{N} = 2$ breaking mass parameter $m$. Thus, *for small $m/\Lambda$ ratios, the domain wall is thick and the field profiles inside it are governed by the Seiberg–Witten theory; but for large $m/\Lambda$, the domain wall is thin and the SW theory simply does not apply.*

## 6. Gaugino Condensates and Effective Superpotentials

When the scalar field $\Phi$ is light, its condensate $U = \langle \operatorname{tr} \Phi^2 \rangle$ dominates the low-energy dynamics of the softly broken $\mathcal{N} = 2$ SSYM theory, hence the success of the SW theory whose primary focus is on the $U$ dependence of various quantities. The low-energy importance of $\Phi$ and $U$ diminishes with $\Phi$'s mass $m$; in the large $m$ limit, $\Phi$ and one of the two gauginos decouple leaving us with an effective $\mathcal{N} = 1$ SSYM theory with

$$\Lambda_{\text{eff}} \;\equiv\; \Lambda_1 \;=\; \sqrt[3]{-m\Lambda_2^2}. \tag{6.1}$$

(In this section we refer to the $\Lambda$ parameter of the high-energy $\mathcal{N} = 2$ SSYM

theory as $\Lambda_2$ in order to to distinguish it from the $\Lambda_1$.) This effective $\mathcal{N}=1$ SSYM theory has $N_c$ isolated vacua related to each other by a spontaneously broken $Z_{N_c}$ symmetry; the order parameter distinguishing between the vacua is the *gaugino condensate*[10]

$$S \equiv \tfrac{1}{16\pi^2} \langle \operatorname{tr} \lambda^\alpha \lambda_\alpha \rangle = \Lambda_1^3 \, e^{2\pi i n/N_c}, \quad n = 1, 2, \ldots, N_c. \tag{6.2}$$

From the infrared point of view, $S$ is the scalar component of the composite chiral superfield $\tfrac{1}{16\pi^2} \operatorname{tr}(W^\alpha W_\alpha)$ describing the lightest glueballs/oddballs of the SSYM theory. Integrating out all the heavier particles leaves us with an effective theory for the gaugino condensate $S$. The Veneziano–Yankielowicz superpotential[10]

$$W(S) = S \log \frac{S^{N_c}}{(e\Lambda_1^3)^N} \tag{6.3}$$

of this theory is exact — it is completely determined by the R-anomaly of the SSYM — but multi-valued (the logarithm is only defined modulo $2\pi i$); altogether, it has $N_c$ supersymmetric vacua (6.2).

For our purpose of constructing BPS domain wall solutions, the Veneziano–Yankielowicz–Taylor[11] effective theory of the gaugino condensate $S$ has two major deficiencies. First, as argued by Kovner and Shifman,[1] different vacua (6.2) belong to different branches of the multi-valued superpotential (6.3), which means that additional, heavier degrees of freedom become excited in the mid-wall region of space where $W_{\text{eff}}(S(x_3))$ jumps from one branch of (6.3) to another. Without somehow accounting for such heavier degrees of freedom we would have an apparent discontinuity of $W(x_3)$ and hence an unresolved singularity of the energy density of a BPS domain wall

$$\frac{d\,\text{Energy}}{d\,\text{Volume}} = 2 \left| \frac{dW}{dx_3} \right|. \tag{6.4}$$

This problem becomes particularly severe in the large $N_c$ limit where this singularity accounts for almost all of the wall's energy; this problem is discussed in great detail in refs. [3] and we have nothing to add to *that* discussion in this article.

On the other hand, for $N_c = 2$ the superpotential $W(S(x_3))$ is actually continuous: Indeed, both vacua $S = -\Lambda_1^3$ and $S = +\Lambda_1^3$ of the $SU(2)$ SSYM are invariant under the charge conjugation symmetry of the theory,[*] $S \to S^*$. The entire domain wall solution is therefore C-invariant, *i.e.*, real $S(x_3)$ throughout the wall; the superpotential $W(S(x_3))$ is given by the real branch of (6.3) for real $S > 0$ for $x_3 < 0$ and by a different branch that is real for real $S < 0$ for $x_3 > 0$, but both branches have $W = 0$ for $S = 0$ at $x_3 = 0$. Thus we have no discontinuity and the two branches of the $W(S(x_3))$ account for the entire energy of the BPS domain wall.

The other major difficulty with the Veneziano–Yankielowicz–Taylor[11] theory is that it does not provide us with an effective Kähler function for the gaugino condensate, never mind the higher-derivative terms in the effective low-energy Lagrangian. From the SSYM point of view, even the Kähler term $\int d^4\theta\, K(S, \overline{S})$ is a high-derivative term in the gauge theory's generating functional $\Gamma$ and hence subject to all kinds of perturbative and non-perturbative corrections. Thus, the derivative interactions of the $S$ superfield remain quite unknown[†] and because of this lack of crucial data, we have no way of calculating the actual profile $S(x_3)$ of the BPS domain wall configuration, not even for $N_c = 2$. All we can say with confidence is that the overall thickness $w$ of the wall should be of the order $O(1/|\Lambda_1|)$ (give or take a power of the $N_c$), simply because it's the only distance scale of the $\mathcal{N} = 1$ SSYM theory.

---

[*] Without loss of generality, we assume real $\Lambda_1^3 > 0$.
[†] Qualitatively, we expect the Kähler metric $g_{S\overline{S}}$ to be non-singular — otherwise, contrary to Witten's supersymmetric index theorem, there would be extra supersymmetric vacua in addition to (6.2). (Actually, a chirally invariant vacuum with a massless fermion at $S = 0$ would be consistent with the Witten's index, but contrary to some statements in the literature,[1–4] there are no physical reasons for the existence of such a vacuum.) Unfortunately, Kähler functions of $\mathcal{N} = 1$ theories are not related to holomorphic functions (such as prepotentials of $\mathcal{N} = 2$ theories), so absence of singularities is not much of a constraint.

\* \* \*

Let us now turn our attention to the regime of $m \sim \Lambda_2 \sim \Lambda_1$ when both gaugino condensate and scalar condensates are equally important. For $N_c = 2$, the only gauge-invariant condensate of the adjoint field $\Phi$ is $U = \langle \operatorname{tr} \Phi^2 \rangle$ and the Veneziano–Yankielowicz–Taylor superpotential[11] for the $S$ and $U$ condensates is completely determined by the global $U(1)_A \times U(1)_R$ symmetries of the $\mathcal{N}=1$ gauge theory and their anomalies;[19,20]

$$W(S, U) = S \log \frac{U^2}{\Lambda_2^4} + mU. \tag{6.5}$$

Again, this superpotential has infinite number of branches but only two physically distinct supersymmetric vacua:

$$\begin{aligned}
1)\quad & U = +\Lambda_2^2, \quad S = -m\Lambda_2^2 = +\Lambda_1^3, \quad W = +m\Lambda_2^2, \\
2)\quad & U = -\Lambda_2^2, \quad S = +m\Lambda_2^2 = -\Lambda_1^3, \quad W = -m\Lambda_2^2.
\end{aligned} \tag{6.6}$$

The superpotential (6.5) is exact, thus the two vacua (6.6) persist for all non-zero values of the mass parameter $m$. Indeed, for small $m \ll \Lambda_2$, the two vacua (2.7) and (2.9) of the Seiberg–Witten theory are in exact agreement with the eqs. (6.6); likewise, the two vacua (6.2) of the effective pure-gauge $SU(2)$ SSYM theory for large $m \gg \Lambda_2$ also agree with the (6.6).

Unfortunately, the Veneziano–Yankielowicz–Taylor theory does not provide us with a Kähler function $K(S, U, \overline{S}, \overline{U})$ or any knowledge of the higher-derivative interactions of the $S$ and $U$ superfields. Consequently, we are unable to calculate the field profiles $S(x_3)$ and $U(x_3)$ for the domain wall solution interpolating between the two vacua (6.6). Nevertheless, we can make a qualitative statement: *There should be BPS-saturated domain wall solutions for all $m \neq 0$.* Indeed, we know that such solutions do exist in the Seiberg–Witten regime of small $m \ll \Lambda_2$, *cf.* section 4. We do not know what exactly happens in the $m \gtrsim \Lambda_2$ regime, but if a BPS saturated wall were to suddenly disappear at some finite value of the $m/\Lambda_2$ ratio, it would mean some kind of a discontinuous phase transition. On the other hand, the vacuum structure (6.6) smoothly continues for all $m \ne 0$, which strongly militates against any discontinuities at finite $m/\Lambda_2$. Although this is not a rigorous proof of existence of BPS saturated walls at large $m$ (and hence all $m \ne 0$), it is a very strong argument in their favor.

$$* \quad * \quad *$$

We conclude this section with a discussion of the three gaugino species in the $SU(2)$ gauge theory with an adjoint scalar, namely the electrically neutral 'photino' $\lambda_\alpha^{(0)}$ and the charged 'winos' $\lambda_\alpha^{(\pm)}$. Because of the gauge symmetry, the distinction between the gaugino species arises from the scalar field: For example, the photino field can be identified in gauge invariant terms as

$$\lambda_\alpha^{(0)} \stackrel{\text{def}}{=} \frac{\text{tr}(\lambda_\alpha \Phi)}{\sqrt{U}}. \tag{6.7}$$

Notice that according to this definition, $\lambda_0^\alpha$ is the lowest component of a gauge-invariant chiral superfield $\text{tr}(W^\alpha \Phi)/\sqrt{U}$, which belongs to the chiral ring of the $\mathcal{N}=1$ theory. Therefore, *in any supersymmetric vacuum of the theory,*

$$\left\langle \lambda^{(0)\alpha} \lambda_\alpha^{(0)} \right\rangle = \left\langle \lambda^{(0)\alpha} \right\rangle \left\langle \lambda_\alpha^{(0)} \right\rangle \equiv 0. \tag{6.8}$$

Thus, *the photinos do not form a condensate* and the net gaugino condensate $S$ is actually the wino condensate $S^\pm = \frac{1}{16\pi^2} \left\langle \lambda^{(+)\alpha} \lambda_\alpha^{(-)} \right\rangle$.

The theorem (6.8) assumes all four supercharges of the $\mathcal{N}=1$ theory remain unbroken in the vacuum state; it does not apply to BPS-saturated field configurations that leave two of the supercharges unbroken but break the other two. Consequently, no theorem prohibits the *local* photino condensate field $S^0(x) = \frac{1}{32\pi^2} \left\langle \lambda^{(0)\alpha}(x) \lambda_\alpha^{(0)}(x) \right\rangle$ from acquiring non-zero values within a BPS-saturated domain wall. The only general rules are $S^0(x_3) \to 0$ for $x_3 \to \pm \infty$ — in the vacuum domains on either side of the wall, the photino condensate must

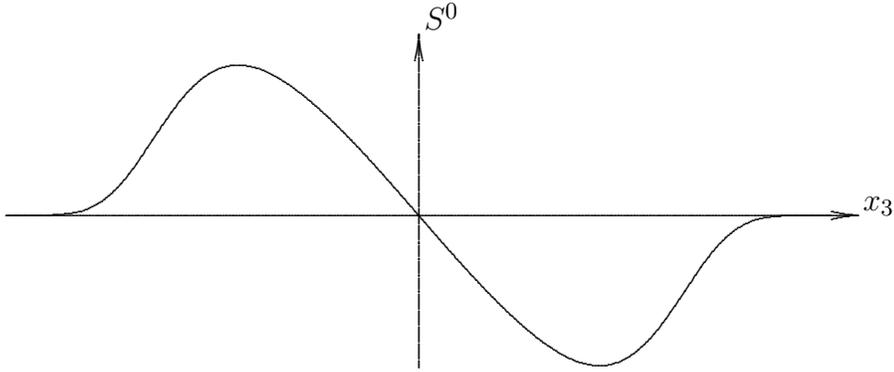

Figure 6: *Qualitative picture of the photino condensate profile inside a BPS domain wall.*

vanish — and $S^0(x_3 = 0) = 0$ following from the $Z_2$ symmetry of the domain wall. Figure 6 illustrates a generic photino-condensate profile $S^0(x_3)$ consistent with these rules.

Of course, just because a non-zero photino condensate is allowed to exist within a BPS domain wall, no theorem requires its actual presence. In order to show that the photino condensate is indeed present within the wall, let us consider the Seiberg–Witten limit $m \ll \Lambda_2$. In the perturbed Coulomb phase prevalent throughout most of the SW wall, the only light fields are the scalar $U$, its superpartner $\psi$, the photon and the photino $\lambda^{(0)}$, all governed by the effective component-field Lagrangian

$$\begin{aligned}
\mathcal{L} &= g^{U\bar{U}} \left( |\partial_\mu U|^2 + \tfrac{i}{2}\bar\psi \slashed{D}\psi + \left| m + \frac{i}{16\pi}\frac{d\tau}{dU}\lambda^{(0)}\lambda^{(0)} \right|^2 \right) \\
&+ \frac{1}{4\pi \operatorname{Im}\tau} \left( \tfrac{1}{4}F_{\mu\nu}^2 + \tfrac{i}{2}\bar\lambda^{(0)}\slashed{D}\lambda^{(0)} + \tfrac{1}{4}\left|\frac{d\tau}{dU}\lambda^{(0)}\psi\right|^2 \right) \\
&+ \frac{i}{16\pi}\frac{d^2\tau}{dU^2}(\lambda^{(0)}\lambda^{(0)})(\psi\psi) + \text{H. c.}
\end{aligned} \qquad (6.9)$$

Of particular interest to us is the fermionic mass term

$$\mathcal{L}_{\text{mass}} = \frac{im^* g^{U\overline{U}}}{16\pi} \frac{d\tau}{dU} \lambda^{(0)}\lambda^{(0)} + \text{H.c.} = \frac{-i}{16\pi} \frac{d\tau(U)}{dx_3} \lambda^{(0)}\lambda^{(0)} + \text{H.c.} \quad (6.10)$$

(*cf.* eq. (4.1)): Inside a BPS domain wall, the photino has a (position-dependent) Majorana mass. Consequently, at the one-loop level of the effective theory, we have the photino condensate

$$\left\langle \lambda^{(0)}\lambda^{(0)} \right\rangle(x) = \lim_{x' \to x} \langle x | \text{tr(photino propagator)} | x' \rangle \neq 0. \quad (6.11)$$

Unfortunately, the ultraviolet limit here diverges quadratically as $m/(x'-x)^2$. Of course, in the full theory this divergence is somehow cut off at the strong interaction scale $\Lambda_2$, thus

$$S^0(x_3)\Big|_{\text{wall}} = O(m\Lambda_2^2), \quad (6.12)$$

but without detailed knowledge of this effective cutoff we are unable to calculate the actual profile $S^0(x_3)$ of the photino condensate.

Finally, a few words about the winos and the wino condensate $S^\pm(x_3)$. In the massless Seiberg–Witten theory, the winos belong to short vector multiplets $W^\pm$ of the $\mathcal{N} = 2$ supersymmetry and their physical mass saturates the BPS limit, $M_W = 2|A_1(U)|$, *cf.* eq. (2.2). As argued in refs. [13], there is a circle-like line in the complex $U$ plane defined by the marginal-stability condition

$$\text{Im}\big(A_1^*(U) A_2(U)\big) = 0. \quad (6.13)$$

Outside this line (*i.e.*, for sufficiently large $U$), the $W^\pm$ multiplets are stable, but they become marginally unstable along the line (6.13) and disappear from the particle spectrum for small $U$ inside the line. In the field-theoretical terms, the kinetic energies of the $W^\pm$ fields become negative for the values of $U$ inside the line (6.13). In a massive SW theory, there are corrections to kinetic energies of various fields due to the $\mathcal{N} = 2$ breaking mass $m$; thus we expect $O(m)$ corrections to the location of the marginal stability line for the $W^\pm$ vectors and winos. Indeed, the two vacua (2.7) and (2.9) of the massive SW theory lie directly atop the massless marginal stability line (6.13), but the winos are actually stable enough to form the wino condensates $S^\pm = \pm m\Lambda_2^2$ (*cf.* eq. (6.6)). Altogether, for small but non-zero masses $m$, the stability region of the winos should look like the dark-grey area on figure 7.

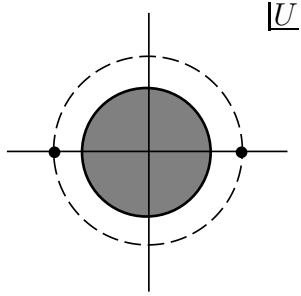

Figure 7: *Schematic picture of the winos' stability region in the complex $U$ plane. The dashed line here represents the marginal stability line (6.13) of the massless SW theory: Winos are stable outside this line but do not exist inside it. For the SW theory with a small $m \neq 0$, the region where winos exist is colored white while the region where they are absent from the particle spectrum is colored dark grey. The two •'s indicate the stable vacua $U = \pm\Lambda_2^2$ of the massive theory.*

Now consider a SW domain wall where the $U$ field smoothly changes from $+\Lambda_2^2$ to $-\Lambda_2^2$. Through most of the wall's interior region, $U(x_3)$ traverses the dark grey region of figure 7 where the particle spectrum of the theory does not contain the $W^\pm$ vector bosons and their wino superpartners. For small $m$, we may neglect the effects of a small gradient of the $U$ field on the wino kinetic energy, thus *through most of the domain wall's interior, winos do not exist and there is no wino condensate.* However, near either end of the domain wall, $U(x_3)$ crosses into the white region of figure 7 where the winos do exist and presumably do form a condensate. Indeed, the wino condensate $S^\pm(x_3)$ must exist and asymptote to $\pm m\Lambda_2^2$ for $x \to \pm\infty$ since these are the vacuum values of the net gaugino condensate $S = S^\pm + S^0$ while the photino condensate $S^0$ has to asymptote to zero as one goes into the domain of a stable $\mathcal{N} = 1$ vacuum, *cf.* eq. (6.8) and figure 6. To summarize, figure 8 shows the qualitative behavior of the wino and photino condensates in the Seiberg–Witten regime ($m \ll \Lambda_2$) of the BPS domain wall.

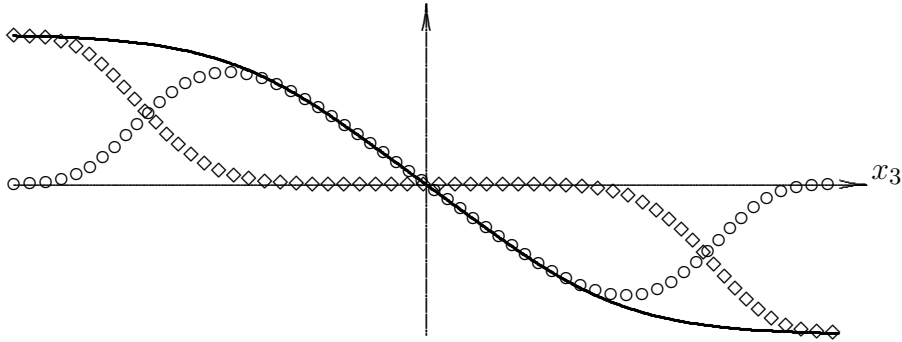

Figure 8: *Qualitative behavior of the gaugino condensates in a Seiberg–Witten domain wall. The circles here represent the photino condensate $S^0(x_3)$, the diamonds — the wino condensate $S^\pm(x_3)$ and the solid line shows the net gaugino condensate $S(x_3) = S^0 + S^\pm$.*

In the large $m \gtrsim \Lambda_2$ regime of the BPS domain wall, the gaugino condensates behave quite differently. In the extreme $m \gg \Lambda_2$ — and hence $m \gg \Lambda_1 \gg \Lambda_2$ — regime, the scalar field $\Phi$ decouples, its expectation value becomes unimportant and the distinction between the photino and the winos becomes unphysical. (Formally, as long as $U \neq 0$, we may define a "photino" field according to eq. (6.7), but the physical meaning of such definition becomes rather obscure when $|U| \ll \Lambda_1$). Consequently, only the net gaugino condensate $S(x_3)$ is meaningful in this regime.

On the other hand, for $m \sim \Lambda_2$ we can readily distinguish the photinos from the winos. At the same time however, there are large $m$-dependent corrections to the SW formulæ for the kinetic energies of the $W^\pm$ and wino fields, so it is quite possible that these fields exist throughout the domain wall, including the $U \approx 0$ region in the wall's middle. If this is indeed the case, we expect both the photino and the wino condensates to be present throughout the wall but to follow different profiles, $S^\pm(x_3) \neq S^0(x_3)$. As of this writing, our knowledge of this regime is limited to generalia; this is a subject for future research.

## 7. BPS Domain walls in MQCD

In this last section of the paper, we shift our focus from the supersymmetric gauge theory to another theory in the same universality class, namely the MQCD.[14] Generally, MQCD is a geometric theory of the M5-brane of the M-theory living in space of geometry $\mathbf{R}^{1,3} \otimes \mathcal{Y}^7$ and completely spanning the ordinary four-dimensional spacetime $\mathbf{R}^{1,3}$. For the vacuum states of MQCD, the M5-brane geometry[16,1521] is $\mathbf{R}^{1,3} \otimes \Sigma$ for some Riemann surface $\Sigma \subset \mathcal{Y}^7$; different vacua have different Riemann surfaces $\Sigma' \neq \Sigma$. The non-vacuum solutions have $M5 = \mathbf{R}^{1,3} \times \Sigma(x_0, x_1, x_2, x_3)$; in particular, a domain-wall solution has $x_3$-dependent $\Sigma(x_3)$ interpolating between two vacua $\Sigma_1$ and $\Sigma_2$. In other words,

$$M5 = \mathbf{R}^{1,2} \otimes \mathcal{S}, \tag{7.1}$$

where $\mathcal{S} \subset \mathbf{R}_{x_3} \otimes \mathcal{Y}^7$ is a three-cycle with boundaries $\Sigma_1$ at $x_3 = -\infty$ and $\Sigma_2$ at $x_3 = +\infty$.

Supersymmetry requires $\mathcal{Y}^7 = \mathcal{Y}^6 \otimes \mathbf{R}$ where $\mathcal{Y}^6$ is a Kähler space; for the problem at hand, $\mathcal{Y}^6 = \mathbf{R}^5 \otimes S^1$ is actually flat. The supersymmetric vacua correspond to the supersymmetric 2–cycles of the $\mathcal{Y}^6$ *i.e.*, the Riemann surface $\Sigma$ should be holomorphically embedded into the $\mathcal{Y}^6$. Likewise, BPS-saturated domain walls that preserve two unbroken supercharges correspond to the supersymmetric 3–cycles *i.e.*, associative submanifolds $\mathcal{S} \subset \mathbf{R}_{x_3} \otimes \mathcal{Y}^6$.[22,23] The associativity condition is best explained[24,25] in terms of a $G_2$ structure of the $\mathbf{R}_{x_3} \otimes \mathcal{Y}^6$ manifold, namely the invariant 3–form

$$\phi = \tfrac{i}{2} dx_3 \wedge (dz_1 \wedge d\bar{z}_1 + dz_2 \wedge d\bar{z}_2 + dz_3 \wedge d\bar{z}_3) + \mathrm{Im}\,(dz_1 \wedge dz_2 \wedge dz_3) \quad (7.2)$$

where $(z_1, z_2, z_3)$ — collectively denoted $\vec{z}$ — are the complex coordinates of the $\mathcal{Y}^6$. The volume form (for the induced Riemannian metric) of any 3–cycle of such a manifold is limited from below by the pull-back of the 3–form (7.2); the associative 3–cycles saturate this lower bound and hence have lowest volumes allowed by their topologies and boundary conditions. (This is similar to the holomorphic 2–cycles of a Kähler manifold having lowest areas allowed for their topologies and boundary conditions.)

Specifically, let $X_1$, $X_2$ and $X_3$ be generic world-volume coordinates of the 3–cycle $\mathcal{S}$ and let $\alpha_i = dx_3/dX_i$ and $\vec{a}_i = d\vec{z}/dX_i$. In terms of these derivatives, the pull-back of the 3–form (7.2) is

$$\phi_{\mathrm{pb}} = \hat{\phi}\, dX^1 \wedge dX^2 \wedge dX^3 \quad \text{where } \hat{\phi} = \mathrm{Re}\,[(\vec{a}_1 \times \vec{a}_3) \cdot \vec{a}_3] + \tfrac{1}{2}\epsilon_{ijk}\,\alpha_i\,\mathrm{Im}\,[\vec{a}_j^* \cdot \vec{a}_k] \quad (7.3)$$

while the volume form is of course $V = \sqrt{\det(h)}\, dX^1 \wedge dX^2 \wedge dX^3$ for the induced metric

$$h_{ij} = \alpha_i \alpha_j + \mathrm{Re}\,[\vec{a}_i^* \cdot \vec{a}_j]. \quad (7.4)$$

Given these formulæ, straightforward but tedious algebra yields

$$\det(h_{ij}) - \hat{\phi}^2 = \left|\vec{R}\right|^2 + \mathrm{Im}^2\,[(\vec{a}_1 \times \vec{a}_3) \cdot \vec{a}_3] \quad (7.5)$$

where

$$\vec{R} = \tfrac{1}{2}\epsilon_{ijk}\,[\alpha_i\,(\vec{a}_j \times \vec{a}_k) - \vec{a}_i^*\,(\vec{a}_j^* \cdot \vec{a}_k)]\,; \quad (7.6)$$

furthermore, $\vec{R} = 0$ implies $\mathrm{Im}\,[(\vec{a}_1 \times \vec{a}_3) \cdot \vec{a}_3] = 0$. Therefore, *a 3–cycle is associative and the domain wall is BPS saturated if and only if* $\vec{R} = 0$.

In order to make practical use of this theorem, let us choose the world-volume coordinates according to

$$X^1 + iX^2 = z^3, \qquad X^3 = x^3 \tag{7.7}$$

which gives us

$$\begin{aligned}
\alpha_1 &= 0, & \vec{a}_1 &= (\partial_1 z_1, \partial_1 z_2, 1), \\
\alpha_2 &= 0, & \vec{a}_2 &= (\partial_2 z_1, \partial_2 z_2, i), \\
\alpha_3 &= 1, & \vec{a}_3 &= (\partial_3 z_1, \partial_3 z_2, 0).
\end{aligned} \tag{7.8}$$

In light of these formulæ, the $\vec{R} = 0$ equation can be solved as

$$\vec{a}_3 = \frac{(\vec{a}_1 \times \vec{a}_2)^* + c_1 \vec{a}_1 + c_2 \vec{a}_2}{\mathrm{Im}[\vec{a}_1^* \cdot \vec{a}_2]} \tag{7.9}$$

for some real coefficients $c_1$ and $c_2$ whose explicit values follow from the consistency of eq. (7.9) with the eqs. (7.8):

$$c_1 = -\mathrm{Re}(J), \quad c_2 = +\mathrm{Im}(J), \qquad \text{where} \quad J = \partial_1 z_1 \, \partial_2 z_2 - \partial_1 z_2 \, \partial_2 z_1. \tag{7.10}$$

Re-phrasing eq. (7.9) as a differential equation for the independent coordinates $z_{1,2}(X)$ of a BPS domain wall, we arrive at

$$\begin{aligned}
F \, \partial_3 z_1 &= (c_1 \partial_1 + c_2 \partial_2) z_1 - (i\partial_1 + \partial_2) z_2^*, \\
F \, \partial_3 z_2 &= (c_1 \partial_1 + c_2 \partial_2) z_2 + (i\partial_1 + \partial_2) z_1^*,
\end{aligned} \tag{7.11}$$

where $c_1$ and $c_2$ are as in eq. (7.10) and

$$F \equiv \mathrm{Im}[\vec{a}_1^* \cdot \vec{a}_2] = 1 + \mathrm{Im}[\partial_1 z_1^* \, \partial_2 z_1 + \partial_1 z_2^* \, \partial_2 z_2]. \tag{7.12}$$

Note that eqs.(7.11) *disallow* holomorphic embedding of any fixed–$x_3$–slices $\Sigma(x_3)$ of the BPS domain wall — except asymptotically for $x_3 \to \pm\infty$ where $\Sigma(x_3)$ asymptotes to holomorphic 2–cycles $\Sigma_{1,2}$. Indeed, suppose there were a holomorphic slice in which $z_1$ and $z_2$ are holomorphic functions of the $z_3 = X_1 + iX_2$. In that case, eq. (7.10) would yield $c_1 = c_2 = 0$ while at the same time $(i\partial_1 + \partial_2)z_{1,2}^* = 0$; consequently, eqs. (7.11) would immediately result in $\partial_3 z_1 = \partial_3 z_2 = 0$. And since eqs. (7.11) are first-order differential equations with respect to the $\partial_3$, that would actually mean complete $X_3 = x_3$ independence of the $z_1$ and $z_2$ — and the solution would be an $x$–independent vacuum, not a domain wall.

Conversely, a constant vacuum solution should have holomorphically embedded slices $\Sigma(x_3) \equiv \Sigma$. Ultimately, this is related to four rather than two supercharges being unbroken, but we can also see how this degenerate case works in context of the $\vec{R} = 0$ equation: For the $x_3$–independent $z_1$ and $z_2$, the $\vec{R}$ reduces to $\vec{a}_1 \times \vec{a}_2$, so $\vec{R} = 0$ implies $\vec{a}_1 \propto a_2$; consequently $\vec{a}_2 = i\vec{a}_1$ (cf. eqs. (7.8)) i.e., $\partial_2 \vec{z} = i\partial_1 \vec{z}$ and hence $\vec{z} = \vec{z}(X_1 + iX_2)$ — a holomorphic embedding.

Now consider the supersymmetric vacua $\Sigma_{1,2}$ that serve as boundaries of the BPS wall's 3–cycle $\mathcal{S}$ — and hence as boundary conditions for the equations (7.11). For the scalar-less $\mathcal{N} = 1$ SSYM theory, the $N_c$ vacua (6.2) correspond to Riemann surfaces

$$z_1^{N_c} = e^{-z_3}, \qquad z_2^{N_c} = e^{+z_3}, \qquad z_1 z_2 = S_n, \quad n = 1, \ldots, N_c \qquad (7.13)$$

where $S_n = e^{2\pi i n/N_c}$ is precisely the gaugino condensate in $\Lambda_1^3$ units.[16,21] Note that the $z_3$ coordinate is periodic modulo $2\pi i$, so eq. (7.13) describes exactly $N_c$ Riemann surfaces distinguishable by the $Z_{N_c}$ phase of the gaugino condensate. Moreover, each of these surfaces $\Sigma_n$ is invariant under a global $U(1)$ symmetry

$$z_1 \mapsto e^{i\delta} z_1, \qquad z_2 \mapsto e^{-i\delta} z_2, \qquad z_3 \mapsto z_3 + iN_c\delta. \qquad (7.14)$$

As argued by Witten,[16] this exact symmetry of MQCD is completely invisible in the ordinary Supersymmetric QCD, and hence cannot be spontaneously broken by a domain wall solution: Indeed, any such breaking would result in some kind of a wall-bound Goldstone mode, which would not make any sense at all from the ordinary S-QCD point of view. Therefore, the entire domain-wall's 3–cycle $\mathcal{S}$ has to be invariant under the symmetry (7.14), which allows us to introduce the following ansatz:

$$\begin{aligned} z_1(X) &= P(X_1, X_3)\, e^{iX_2/N_c}, \\ z_2(X) &= Q(X_1, X_3)\, e^{-iX_2/N_c}, \\ z_3(X) &= X_1 + iX_2 \end{aligned} \quad (7.15)$$

for some complex functions $P(X_1, X_3)$ and $Q(X_1, X_3)$. In terms of these functions, eqs. (7.11) become

$$\begin{aligned} F\,\partial_3 P &= \left(c_1 \partial_1 + \tfrac{i}{N_c} c_2\right) P - i\left(\partial_1 + \tfrac{1}{N_c}\right) Q^*, \\ F\,\partial_3 Q &= \left(c_1 \partial_1 - \tfrac{i}{N_c} c_2\right) Q + i\left(\partial_1 - \tfrac{1}{N_c}\right) P^*, \end{aligned} \quad (7.16)$$

where

$$J = \tfrac{-i}{N_c} \partial_1(PQ) \quad \text{and} \quad F = 1 + \tfrac{1}{2N_c} \partial_1(|P|^2 - |Q|^2) \quad (7.17)$$

while the boundary conditions are simply

$$\begin{aligned} P &= e^{X_1/N_c}, \quad Q = S_1\, e^{-X_1/N_c} \quad \text{for} \quad X_3 \to -\infty, \\ P &= e^{X_1/N_c}, \quad Q = S_2\, e^{-X_1/N_c} \quad \text{for} \quad X_3 \to +\infty. \end{aligned} \quad (7.18)$$

Unfortunately, eqs. (7.16) do not have any non-trivial solutions for which $S(X) \equiv z_1 z_2 = PQ$ is a function of the $X_3$ but not of the $X_1$; we prove this sad theorem in the Appendix B. Thus, the domain wall of the MQCD cannot be interpreted in terms of the $x_3$–dependent gaugino condensate $S(X_3)$. Of course, even for the SSYM theory we could not explain the BPS domain wall in terms of the gaugino condensate alone or calculate the actual profile of the $S(x_3)$, but this was basically a calculational difficulty. In case of the MQCD, the gaugino condensate profile is not even definable, let alone computable even in principle.

* * *

For $N_c > 2$, the boundary conditions (7.18) require $P$ and $Q$ to be inherently complex functions of the $X_1$ and the $X_3$, but in the $SU(2)$ case both the boundary conditions and the equations (7.16) allow for real or purely imaginary $P$ and $Q$. Clearly, this is the MQCD counterpart of having real $S(x_3)$ and $U(x_3)$ for the $SU(2)$ SSYM theory, so let us see how the MQCD equations simplify for the real solutions; specifically, let us assume real $P$ and purely imaginary $Q$. It turns out that in this case, eqs. (7.16) take a simplified form in terms of new real variables $S = -iPQ$ and $R = \frac{1}{2}(P^2 + Q^2)$:

$$\begin{aligned} \partial_3 R &= -\partial_1 S, \\ \left(1 + \tfrac{1}{2}\partial_1 R\right) \partial_3 S &= -\tfrac{1}{2}(\partial_1 S)^2 + \partial_1 R \pm \sqrt{S^2 + R^2}, \end{aligned} \qquad (7.19)$$

while the boundary conditions read

$$S(X_3 \to \mp\infty) \to \pm 1, \qquad R(X_3 \to \mp\infty) \to \sinh(X_1). \qquad (7.20)$$

The first equation (7.19) allows us to introduce a single variable $\psi(X_1, X_3)$ such that $S = \partial_3 \psi$ and $R = -\partial_1 \Psi$. Substituting this into the second eq. (7.19) gives us a *Monge–Ampere* equation

$$\tfrac{1}{2}\left[(\partial_1 \partial_3 \Psi)^2 - (\partial_1^2 \Psi)(\partial_3^2 \Psi)\right] = \nabla^2 \psi \pm \sqrt{(\nabla \psi)^2} \qquad (7.21)$$

subject to the boundary conditions

$$\psi(X_3 \to \mp\infty) = \pm X_3 - \cosh(X_1). \qquad (7.22)$$

This reduction to a single — but fourth order — Monge–Ampere equation was first noticed by A. Volovich.[17] She has also constructed a formal solution as an infinite power series in $(1/\cosh^2 X_3)$; unfortunately, the series diverges, so it is not clear whether her solution actually exists.

The Monge–Ampere equations have been extensively studied by mathematicians[18] and for many equations of this type, the existence of a unique solution is guaranteed by a theorem. Unfortunately, such theorems do not apply to the specific Monge–Ampere eq. (7.21), so at present we do not have a rigorous existence proof for a BPS-saturated domain wall in MQCD. Clearly, more research in this direction would be very helpful.

$$*\quad*\quad*$$

Finally, consider the MQCD corresponding to the supersymmetric gauge theory with an adjoint scalar $\Phi$. In the $SU(2)$ case, the Riemann surfaces $\Sigma$ of supersymmetric vacua have genus $g = 0$ and general analytic form

$$e^{2z_3} + 2e^{z_3}(z_1^2 - U) + \Lambda_2^4 = 0, \qquad z_2^2 - m^2 z_1^2 - 2mS = 0; \quad (7.23)$$

the $g = 0$ requirement allows for just two such surfaces, with parameters $U$ and $S$ precisely as in eqs. (6.6) for the vacua of the gauge theory, and it is these two surfaces

$$\begin{aligned}
\Sigma_1 &: \quad z_1 z_2 = m\left(z_1^2 - e^{z_3} + \Lambda_2^2\right), \quad z_2^2 = m^2\left(z_1^2 - 2\Lambda_2^2\right), \\
\Sigma_2 &: \quad z_1 z_2 = m\left(z_1^2 - e^{z_3} - \Lambda_2^2\right), \quad z_2^2 = m^2\left(z_1^2 + 2\Lambda_2^2\right)
\end{aligned} \quad (7.24)$$

that serve as boundaries of the supersymmetric 3–cycle $\mathcal{S}$ describing the BPS domain wall.

We are particularly interested in the MQCD analogue of the Seiberg–Witten domain wall in the small $m$ limit — where all fields change but slowly with the $x_3$ and the physics inside the wall is in the Coulomb phase of the massless SW theory. By analogy, we expect approximately holomorphic slices $\Sigma(x_3)$, slowly changing with the $x_3$ and also $|z_2| \ll |z_1|$ (in the massless limit, $z_2 \mapsto 0$). Let us therefore assume holomorphic dependence of the $z_1$ coordinate on the $X_1 + iX_2 = z_3$ but allow for a non-holomorphic $z_2(z_3, \bar{z}_3, X_3)$.

In terms of the world-volume coordinates $z_3$, $\bar{z}_3$ and $x_3$ and the corresponding derivatives $\partial = \frac{1}{2}\left(\frac{\partial}{\partial X_1} - i\frac{\partial}{\partial X_2}\right)$, $\bar{\partial} = \frac{1}{2}\left(\frac{\partial}{\partial X_1} + i\frac{\partial}{\partial X_2}\right)$ and $\nabla = \frac{\partial}{\partial x_3}$, eqs. (7.11) become

$$\begin{aligned} F\nabla z_1 &= -J^*\partial z_1 - J\bar{\partial} z_1 - 2i\partial z_2^*, \\ F\nabla z_2 &= -J^*\partial z_2 - J\bar{\partial} z_2 + 2i\partial z_1^*, \end{aligned} \tag{7.25}$$

where

$$\begin{aligned} F &= 1 + |\partial z_1|^2 - |\bar{\partial} z_1|^2 + |\partial z_2|^2 - |\bar{\partial} z_2|^2, \\ J &= 2i(\bar{\partial} z_1\, \partial z_2 - \bar{\partial} z_2\, \partial z_1). \end{aligned} \tag{7.26}$$

In the holomorphic $z_1$ approximation, $F \approx 1 + |\partial z_1|^2$ (since $|z_2| \ll |z_1|$), $J = -2i\bar{\partial} z_2 \partial z_1$ and the eqs. (7.25) simplify to

$$\begin{aligned} \nabla z_1 &= -2i\partial z_2^*, \\ \nabla z_2 &= \frac{2i}{F}\left(\partial z_1 (\bar{\partial} z_2)^2 - \bar{\partial} z_1^*\, \partial z_2^*\, \partial z_2\right). \end{aligned} \tag{7.27}$$

Note that the first equation here implies $\bar{\partial}\partial z_2^* = 0$ and hence

$$z_2(z_3, \bar{z}_3, x_3) = f(z_3, x_3) + g^*(\bar{z}_3, x_3) \tag{7.28}$$

where $f$ and $g$ are holomorphic functions of the $z_3$. The boundary conditions for these functions (and also the $z_1(z_3, x_3)$ are

$$\begin{aligned} z_1 &\to \tilde{\Lambda} \cosh\frac{\tilde{z}_3}{2}, & f &\to m\tilde{\Lambda} \sinh\frac{\tilde{z}_3}{2}, & g &\to 0 \quad \text{for } x_3 \to -\infty, \\ z_1 &\to \tilde{\Lambda} \sinh\frac{\tilde{z}_3}{2}, & f &\to m\tilde{\Lambda} \cosh\frac{\tilde{z}_3}{2}, & g &\to 0 \quad \text{for } x_3 \to +\infty, \end{aligned} \tag{7.29}$$

where $\tilde{\Lambda} = i\sqrt{2}\Lambda_2$ and $\tilde{z}_3 = z_3 - \log \Lambda_2^2$.

The second eq. (7.27) can be further simplified by assuming either $|\partial z_1|^2 \gg 1$ or $|\partial z_1|^2 \ll 1$ in units of the eq. (7.23). These assumptions give us respectively $F \approx |\partial z_1|^2$ or $F \approx 1$, and in both cases holomorphy of the $f$ and $g$ functions can be used to derive separate equations for the $\nabla f$ and $\nabla g$. Alas, in both cases the resulting equations are inconsistent with the boundary conditions (7.29), which means that the BPS domain wall in MQCD defies simplification. At best, $|\partial z_1|^2$ cannot be either large or small throughout the supersymmetric 3–cycle $\mathcal{S}$, but it is also possible that the holomorphy approximation $\bar{\partial} z_1 \approx 0$ has to break down in some part of the 3–cycle's world-volume.

## 8. Summary

Let us now summarize our discussion of the BPS-saturated domain wall configurations. Although their tension can be calculated exactly even in a strongly interacting theory such as a confining $\mathcal{N} = 1$ SSYM, their structure and even their very existence depend on the theory's details; indeed, there is no rigorous proof that the domain walls of the $\mathcal{N} = 1$ SSYM are BPS saturated. Nevertheless, our investigation provides good evidence that they are BPS and sheds much new light on their structure. It also reveals some unexpected intricacies associated with domain walls; hopefully, our results and observations will provide a starting point for further study.

Rather than aiming directly at the domain walls of the $\mathcal{N} = 1$ SSYM, our study was focused on two related theories, namely Seiberg–Witten and MQCD. Most of our work was done in the framework of the $\mathcal{N} = 2$ $SU(2)$ Seiberg–Witten theory perturbed by the second-SUSY-breaking mass of the adjoint chiral superfield. The vacuum structure of this theory follows from the effective superpotential (6.5) for the gaugino and scalar condensates; there are precisely two degenerate vacua which smoothly connect in the low-mass limit to the two singularities of the Seiberg–Witten moduli space where the magnetic monopoles or dyons become massless and condense. Likewise, in the large-scalar-mass limit, we have an effective $\mathcal{N} = 1$ SSYM theory with $N_c = 2$ degenerate vacua, and these are precisely the vacua we get. Nowhere in our analysis we find any sign of existence of additional, non-chiral vacua advocated in earlier work[1] or any kind of a phase transition; instead, we find the same two degenerate vacua for all $m \neq 0$.

For small $m \ll \Lambda$, we have not only established the existence of a BPS saturated domain wall solution but actually calculated its field profiles $U(x_3)$, $\mathcal{M}(x_3)$ and $\mathcal{D}(x_3)$. The profiles — depicted on figure 5 — show a five-layer structure of the wall, with two confining-phase domains at the two ends, a perturbed Coulomb phase domain in the middle and two transition regions between these domains. Based upon continuity of the vacuum structure of the theory, we argued that the BPS domain wall solution exists for all $m \neq 0$ and even in the $m \to \infty$ limit of the $\mathcal{N} = 1$ SSYM. The five-layer structure, however, survives only for very small $m \lesssim \Lambda/400$: For larger masses, the two transition regions take over the whole domain wall and soon enough we have both the magnetic monopole condensate and the dyon condensate in the same region of space in the middle of the wall. Clearly this situation cannot be described in terms of an effective local field theory such as Seiberg–Witten; furthermore, the wall becomes so narrow that the higher-derivative terms in the BPS equations become important. Consequently, we could not calculate the wall's field profiles for larger masses; this remains an open problem for future work.

Our analysis has revealed several interesting properties of the gaugino condensates in theories with adjoint scalars. In vacuum states with four unbroken supercharges, the neutral photinos do not condense and the gaugino condensate is solely due the electrically charged winos. However, there is a non-trivial photino condensate in the middle of a BPS domain wall which breaks two more supercharges and leaves only two supercharges unbroken. Furthermore, for small $m$, the charged winos condense outside the wall and near its ends, but deep inside the wall the charged winos do not even exist. For larger $m$, both the photino and the wino condensates exist throughout the domain wall but follow qualitatively different profiles. (Except in the effectively scalar-less $\mathcal{N} = 1$ SSYM theory for very large $m$ where the very distinction between the winos and the photinos becomes unphysical.)

In the course of our investigation into the Seiberg–Witten domain walls, we encountered several surprises. Among the surprising features which we do not fully understand and which deserve future attention are: the amazing accuracy of the approximation (4.3), the stringency of the limit (4.16) for the validity of the low-mass approximation and the Hamiltonian form of the BPS equations (4.7) for the transitional regions; indeed, the physical meaning of the Hamiltonian (4.9) + (4.10) remains completely obscure. In addition, there is plenty of open questions pertaining to the $SU(N_c > 2)$ theories we have not explored in this article, for example, the issue of the superpotential cusps[3] for $N_c > 2$. However, following the work of Douglas and Shenker,[28] we expect significant simplification of the problem in the large $N_c$ limit. We are currently investigating such large-$N_c$ domain walls and have some preliminary results for the extremely low mass, almost $\mathcal{N} = 2$ Seiberg–Witten regime of the $SU(N_c)$ theory. In particular, we find the same five-layer structure of the BPS domain walls as in the $SU(2)$ case and even the Hamiltonian form of the BPS equations in the transition regions. Furthermore, we confirm Witten's result[16] that the wall's tension is proportional to $N_c$ (rather than $N_c^2$ one naively expects from a solitonic solution).

We have also tried — with limited success — to investigate the BPS domain walls in MQCD. We have translated Witten's geometrical picture[14] of such walls into an explicit set of partial differential equations. We also managed to greatly simplify these equations for some special cases. In particular, for the MQCD analogue of the $\mathcal{N} = 1$ $SU(2)$ SSYM theory, we ended up with a single fourth-order Monge–Ampere equation for one real function. Unfortunately, we could not solve this equation for the domain-wall boundary conditions; also the known mathematical theorem assuring that some Monge–Ampere equations always have a solution does not apply to the specific equation and boundary conditions at hand. Nevertheless, by analogy with the Seiberg–Witten theory, the MQCD most likely does have a BPS domain wall solution we simply have not found yet. Thus, our results make a good starting point for future work; when the solution will be eventually found, it is likely to teach us much about the domain walls in both quantum field theory and string/M theory.

Acknowledgements: The authors wish to thank A. Brandhuber, N. Itzhaki and S. Theisen for many fruitful discussions of domain walls in MQCD. An important part of the work was accomplished while V. K. was visiting the ITP at Santa Barbara; many thanks for the hospitality.

# APPENDIX A
## Higher–Derivative Interactions and their Effect on Auxiliary Fields

In order to solve eqs. (5.1) for a BPS-saturated domain wall of a $\mathcal{N} = 1$ supersymmetric effective field theory we need to solve for the auxiliary fields $F^i$ in terms of the physical scalar fields $A^i$. In the context of an effective Lagrangian — or rather the generating functional $\Gamma$ which we assume to be local, — the only terms that remain relevant in the long-distance or low-energy limit are those with at most four superderivatives,

$$\mathcal{L}_{4\nabla} \;=\; -\int d^4\theta \, K(\overline{\Phi}, \Phi) \;+\; \int d^2\theta \left[ W(\Phi) + \tfrac{\tau(\Phi)}{16\pi i} W^\alpha W_\alpha \right] \;+\; \text{H.\,c.} \qquad (\text{A.1})$$

Expanding this effective Lagrangian in component fields, we focus on terms containing the auxiliary fields $F^i$ and disregard the fermionic fields as not germane to the domain wall. Thus,

$$\mathcal{L}_{4\nabla} \;=\; -g_{i\bar{j}} F^i F^{*\bar{j}} \;+\; W_i \, F^i \;+\; W^*_{\bar{i}} \, F^{*\bar{i}} \;+\; \cdots \qquad (\text{A.2})$$

and hence

$$F^i \;=\; g^{i\bar{j}} W^*_{\bar{j}} \,. \qquad (\text{A.3})$$

Unfortunately, eqs. (A.3) is valid only in the long-distance limit where $\mathcal{L} \approx \mathcal{L}_{4\nabla}$. At shorter distances or higher energies, the higher-derivative interaction terms in the effective Lagrangian become relevant too. At the six superderivative level, we have

$$\begin{aligned}\mathcal{L}_{6\nabla} &= \int d^4\theta \left[ L_{ij}(\overline{\Phi},\Phi) \nabla^\alpha \Phi^i \nabla_\alpha \Phi^j + H(\overline{\Phi},\Phi) W^\alpha W_\alpha + \text{H.c.} \right] \\ &= L_{ij} F^i \nabla^\mu \nabla_\mu A^j + L_{(ij,k)} F^i \nabla^\mu A^j \nabla_\mu A^k \\ &\quad + L_{ij,\bar{k}} F^i F^j F^{*\bar{k}} + H_{\bar{i}} F^{*\bar{i}} \left( D^2 + \tfrac{1}{4}(F_{\mu\nu} + \tilde{F}_{\mu\nu})^2 \right) \\ &\quad + \text{H.c.} + \cdots \end{aligned} \quad (A.4)$$

where the '$\cdots$' again stands for the fermionic terms or terms not involving the auxiliary fields $F^i$ and $F^{*\bar{j}}$. The effect of the the six-superderivative interactions (A.4) on the field equations for the auxiliary fields is to replace eqs. (A.3) with messy, non-linear equations:

$$\begin{aligned} g_{\bar{i}j} F^j &= W^*_{\bar{i}} + L^*_{jk,\bar{i}} F^j F^k + 2 L^*_{\bar{i}\bar{j},k} F^{*\bar{j}} F^k + L^*_{(\bar{i}\bar{j},\bar{k})} \nabla^\mu A^{*\bar{j}} \nabla_\mu A^{*\bar{k}} \\ &\quad + L^*_{\bar{i}\bar{j}} \nabla^\mu \nabla_\mu A^{*\bar{j}} + H_{\bar{i}} \left( D^2 + \tfrac{1}{4}(F_{\mu\nu} + \tilde{F}_{\mu\nu})^2 \right). \end{aligned} \quad (A.5)$$

At the eight superderivative level (and beyond), the field equations for the $F^i$ fields become not just non-linear but differential rather than algebraic. As an example of an eight-superderivative terms, consider

$$\int d^4\theta \, \tfrac{1}{16} N_{\bar{i}j}(\overline{\Phi},\Phi) \, \nabla^{\dot\alpha} \nabla_{\dot\alpha} \overline{\Phi}^{\bar{i}} \, \nabla^\alpha \nabla_\alpha \Phi^j \quad (A.6)$$

which expands to

$$\begin{aligned} & N_{\bar{i}j} \nabla^\mu F^{*\bar{i}} \nabla_\mu F^j + N_{\bar{i}j,k\bar{\ell}} F^{*\bar{i}} F^j \left( \nabla^\mu A^k \nabla_\mu A^{*\bar{\ell}} - F^k F^{*\bar{\ell}} \right) \\ &\quad + N_{\bar{i}j,k} F^{*\bar{i}} \left( \nabla^\mu A^k \nabla_\mu F^j - F^k \nabla^\mu \nabla_\mu A^j \right) \\ &\quad + N_{\bar{i}j,\bar{\ell}} F^j \left( \nabla^\mu F^{*\bar{i}} \nabla_\mu A^{*\bar{\ell}} - \nabla^\mu \nabla_\mu A^{*\bar{i}} F^{*\bar{\ell}} \right) + \cdots. \end{aligned} \quad (A.7)$$

Of particular importance is the first term here, which acts as a kinetic-energy Lagrangian term for the auxiliary — or rather formerly auxiliary and now propagating — fields $F^i$ and $F^{*\bar{i}}$. Consequently, the field equations for the $F^i$ now contain derivatives of the $F^i$ themselves as well as of the $A^i$ fields:

$$g_{\bar{i}j}F^j = (\text{A.5}) - \nabla^\mu\left(N_{\bar{i}j}\nabla_\mu F^j + N_{\bar{i}j,\bar{\ell}}F^j\nabla_\mu A^{*\bar{\ell}}\right) + N_{\bar{i}j,k}\nabla^\mu A^k \nabla_\mu F^j$$
$$+ N_{\bar{i}j,k\bar{\ell}}F^j\nabla^\mu A^k\nabla_\mu A^{*\bar{\ell}} - N_{\bar{i}j,k}F^k\nabla^\mu\nabla_\mu A^j - N_{\bar{\ell}j,\bar{i}}F^j\nabla^\mu\nabla_\mu A^{*\bar{\ell}}$$
$$- (N_{\bar{i}j,k\bar{\ell}} + N_{\bar{\ell}j,k\bar{i}})F^j F^k F^{*\bar{\ell}}. \tag{A.8}$$

Of course, (A.6) gives just one example of an eight-superderivative interaction; when all such interactions are taken into account, the field equations for the $F^i$ fields become a really complicated mess. But such details are not important for our argument; what's important that the field equations for the $F^i$ — and hence the BPS equations (5.1) — become very complicated differential equations we do not know how to solve.

# APPENDIX B
# Proof of $X_1$–Dependence
# of the would-be Gaugino Condensate $S$ in MQCD

In this Appendix we prove that all non-trivial solutions of eqs. (7.16) have $X_1$–dependent $S = PQ$ by assuming the contrary — a solution with $S = S(X_3)$ — and showing that this solution must be trivial. We begin by changing variables according to

$$P = \sigma e^{+\gamma}, \qquad Q = \sigma e^{-\gamma} \tag{B.1}$$

where we assume $\sigma = \sqrt{S}$ to depend only on the $X_3$ and not on the $X_1$. Substituting these variables into eqs. (7.16) results in

$$Fe^{+\gamma}\left(\partial_3\sigma + \sigma\partial_3\gamma\right) = 0 - i\sigma^* e^{-\gamma^*}\left(-\partial_1\gamma^* + \tfrac{1}{N_c}\right),$$
$$Fe^{-\gamma}\left(\partial_3\sigma - \sigma\partial_3\gamma\right) = 0 + i\sigma^* e^{+\gamma^*}\left(+\partial_1\gamma^* - \tfrac{1}{N_c}\right), \tag{B.2}$$

where the $(c_1\partial_1 + \tfrac{ic_2}{N_c})$ terms vanish because of $J = \tfrac{-i}{N_c}\partial_1(PQ) = 0$. A linear re-arrangement of eqs. (B.2) gives us

$$F\partial_3\sigma = +i\sigma^*\left(\partial_1\gamma^* - \tfrac{1}{N_c}\right)\cosh(\gamma + \gamma^*),$$
$$F\sigma\partial_3\gamma = -i\sigma^*\left(\partial_1\gamma^* - \tfrac{1}{N_c}\right)\sinh(\gamma + \gamma^*) \quad (B.3)$$

and hence

$$\frac{\partial_3\sigma}{\sigma} = -\frac{\partial_3\gamma}{\tanh(\gamma + \gamma^*)}. \quad (B.4)$$

The last equation is an ordinary differential equation with respect to the $\partial_3$ and its real part can be easily integrated; this yields

$$\sinh(\gamma + \gamma^*) = \frac{h(X_1)}{|\sigma(X_3)|^2} \quad (B.5)$$

for some $X_1$–dependent (but $X_3$–independent) integration "constant" $h$. In light of the boundary conditions (7.18), we should have $h = \sinh\frac{2X_1}{N_c}$, but out argument does not depend on this specific form.

As a consequence of eq. (B.5), the coefficient $F$ in eqs. (7.16) and (B.3) depends only on the $X_1$ and not on the $X_3$. Indeed,

$$F \equiv 1 + \tfrac{1}{2N_c}\partial_1(|P|^2 - |Q|^2) = 1 + \frac{|\sigma|^2}{N_c}\partial_1\sinh(\gamma + \gamma^*) = 1 + \frac{\partial_1 h(X_1)}{N_c}. \quad (B.6)$$

Next, consider the phase of $\partial_1\gamma^* - \tfrac{1}{2N_c}$. According to the first eq. (B.3),

$$\arg\left(\partial_1\gamma^* - \tfrac{1}{N_c}\right) = \arg(\partial_3\sigma/i\sigma^*), \quad (B.7)$$

and since $\sigma$ does not depend on the $X_1$, the phase of the $(\partial_1\gamma^* - \tfrac{1}{N_c})$ should be $X_1$–independent as well. In other words,

$$\operatorname{Im}\partial_1\gamma = G(X_3)\left(\operatorname{Re}\partial_1\gamma - \tfrac{1}{N_c}\right)$$

and

$$\left(\partial_1\gamma^* - \tfrac{1}{N_c}\right) = \tfrac{1}{2}(1 - iG(X_3))\left(\partial_1(\gamma + \gamma^*) - \tfrac{2}{N_c}\right). \quad (B.8)$$

Let us substitute the last formula into the first eq. (B.3). After a little re-arrangement of factors, we obtain

$$\begin{aligned}\frac{\partial_3 \sigma^2(X_3)}{i + G(X_3)} &= \frac{|\sigma|^2}{F}\left[\partial_1 \sinh(\gamma+\gamma^*) - \tfrac{2}{N_c}\cosh(\gamma+\gamma^*)\right] \\ &= \frac{N\partial_1 h(X_1) - 2\sqrt{h^2(X_1)+|\sigma(X_3)|^4}}{N + \partial_1 h(X_1)}\end{aligned} \quad (B.9)$$

where the left hand side should be $X_1$–independent but the right hand side does depend on the $X_1$ for any non-trivial $h(X_1) \neq$ const.

By *reductio ad absurdum*, Q. E. D.